# A METHODOLOGY FOR ASSESSING AGILE SOFTWARE DEVELOPMENT APPROACHES

Shvetha Soundararajan

Research proposal submitted to the faculty of the Virginia Polytechnic Institute and State University in partial fulfillment of the requirements for the degree of

Doctor of Philosophy
in
Computer Science and Applications

Dr. James D. Arthur (Chair)
Dr. Osman Balci
Dr. Steven D. Sheetz
Dr. Todd Stevens
Dr. Eli Tilevich

March 17, 2011
Blacksburg, VA

# A METHODOLOGY FOR ASSESSING AGILE SOFTWARE DEVELOPMENT APPROACHES

Shvetha Soundararajan

ABSTRACT

Agile methods provide an organization or a team the flexibility to adopt a selected subset of principles and practices based on their culture, their values, and the types of systems that they develop. More specifically, every organization or team implements a customized agile method, tailored to better accommodate its needs. However, the extent to which a customized method supports the organizational objectives, or rather the 'goodness' of that method is questionable. Existing agile assessment approaches focus on a comparative analysis, or are limited in scope and application. In this research, we propose a structured, systematic and comprehensive approach to assess the 'goodness' of agile methods. We examine an agile method based on *(1) its **adequacy***, *(2) the **capability** of the organization to support the adopted principles and practices specified by the method,* and *(3) the method's **effectiveness***. We propose the **O**bjectives, **P**rinciples and **P**ractices (OPP) Framework to guide our assessment. The Framework identifies (1) *objectives* of the agile philosophy, (2) *principles* that support the objectives, (3) *practices* that are reflective of the principles, (4) the *linkages* between the objectives, principles and practices, and (5) *indicators* for each practice to assess the effectiveness of the practice and the extent to which the organization supports its implementation. In this document, we discuss our solution approach, preliminary results, and future work.

# Table of Contents

## List of Figures

## List of Tables

# 1. Introduction

Agile methods are designed for customization; they offer an organization or a team the flexibility to adopt a set of principles and practices based on their culture and values. While that flexibility is consistent with the agile philosophy, it can lead to the adoption of principles and practices that can be sub-optimal relative to the desired objectives. We question then, how can one determine if adopted practices are ‘in sync’ with the identified principles, and to what extent those principles support organizational objectives? In this research, we focus on assessing the ‘goodness’ of an agile method adopted by an organization based on *(1) its adequacy, (2) the capability of the organization to provide the supporting environment to successfully implement the method,* and *(3) the method’s effectiveness*.

To guide our assessment, we propose the Objectives, Principles and Practices (OPP) framework. The design of the OPP framework revolves around the identification of the agile objectives, principles that support the achievement of those objectives, and practices that reflect the ‘spirit’ of those principles. Well-defined linkages between the objectives and principles, and between the principles and practices are also established to support the assessment process. We assess the *adequacy* of an agile method by traversing the linkages in a top-down fashion. That is, given the set of objectives espoused by the agile method, we follow the linkages downward to ensure that the appropriate principles are enunciated, and that the proper practices are expressed. We assess the *capability* of an organization to implement its adopted method and the *effectiveness* of that implementation by using both a top-down and bottom-up traversal of the linkages. The bottom-up assessment, however, is predicated on the identification of people, process, project and product properties associated with each practice that attest to the presence and implementation of that practice. We refer to each practice, property pair as an indicator. By following the linkages upward from the indicators, we can infer the use of proper principles and the achievement of desired objectives. This document outlines our proposed research, preliminary efforts and future directions.

## *1.1 Motivation*

Currently, the number of organizations adopting agile methods is increasing - some of the reasons being *(1) the ability to accommodate change throughout the development lifecycle, (2) improved quality, (3) greater return on investment, (4) shorter development periods, (5) improved customer satisfaction, (6) better team morale, (7) reduced waste* and *(8) better predictability* [1-3]. Agile adoption in an organization is guided primarily by its culture, values and the types of systems being developed by the organization. More specifically, when an organization decides to adopt an agile method, we ask the following questions:

1. Does the agile method adopted have the potential to satisfy the values (various goals) of the organization? More specifically, does the method have the principles and practices in place to achieve the touted values?

2. Does the culture of the organization allow the adoption and application of the agile method? Does the organization's environment have the ability to support the implementation of the method? For example, if the people of an organization are resistant to change, getting them to adopt agile methods is a difficult undertaking that may be futile.

3. Is the agile method suited to handle the types of systems (small-, medium-, and larger-scale, mission and life-critical, etc.) that are being developed by the organization? For example, can the method support the development of a larger scale system that would be built over the period of three years?

An organization's objectives are reflective of its culture and values. However, how can one determine if the adopted agile method is consistent with its objectives?

The agile philosophy provides an organization or a team the flexibility to adopt a selected subset of principles and practices. However, more often than not, these customized approaches do not reflect the agile principles associated with the practices. Also, organizations lack the supporting environment to effectively implement the adopted methods. As a result, the benefits afforded by agile methods are not fully realized [4]. Hence, once again, we question the extent to which an agile method or a customized approach satisfies the needs of an organization. In effect, we question the 'goodness' of that approach.

The agile philosophy places utmost importance on "working software" being the primary measure of progress. Hence, most assessment approaches for agile methods focus on assessing the working software and process artifacts. In particular, they place emphasis on the *product* and predominantly ignore potential measures reflecting process, project, and people characteristics. Nonetheless, there are selected approaches that assess the *process*. There exist Agile Process Improvement Frameworks such as the Sidky Agile Measurement Index (SAMI) [5, 6] and Agile Adoption and Improvement Model (AAIM) [7] that guide an organization's agile adoption and improvement efforts. Both frameworks describe levels of agility modeled on similar concepts found in the Software Capability Maturity Model (SW CMM) [8] and Capability Maturity Model Integration (CMMI) [9]. That is, a set of practices is to be adopted by an organization at each level in order to be 'agile' at that level. The primary disadvantage of these frameworks is that a set of practices is 'forced' on an organization at defined levels, which compromises the flexibility offered by agile methods. Third party agility measurement tools such as Comparative Agility [10, 11] and the Thoughtworks Agile Assessment survey [12, 13] focus on assessing the extent to which an organization or a team is successful in adopting and using agile methods. These tools help determine the presence or absence of practices in an organization rather than the degree to which those practices are used. Hence, there is a need for a more comprehensive approach to assessing the 'goodness' of agile methods.

## *1.2 Problem Statement*

An agile method is composed of a set of objectives that the method proposes to achieve, principles that support those objectives, and practices that are reflective of the stated principles. An organization's culture (people and their attitudes), its values (beliefs and ideas about what kind of goals should be pursued), and the types of systems being developed (small-, medium- or larger-scale, mission and life-critical, network-centric, etc.), define the composition of an adopted agile method. Given the various factors that shape an agile method within an organization, how can we determine if that method is 'in-sync' with the organizational objectives? ***In short, how can we assess the 'goodness' of an agile method?***

In assessing the 'goodness' of an agile method, we examine the method from three perspectives:

1. *We review the composition of the agile method independent of any organizational objectives.*

   The agile community has endorsed many standalone agile methods such as eXtreme Programming, Feature Driven Development, Crystal, Lean, etc. How can organizations intending to adopt any of those accepted methods, or attempting to design a customized approach, ensure that the method is sufficient with respect to meeting its stated objectives?

2. *We study an instance of the agile method defined by the organization.*

   An instance of an adopted agile method should reflect the culture and values of the organization. Hence, the objectives, principles, and practices touted by the instance may differ from the composition of the method in its original form.

   Does the instance state desired objectives that reflect the organizational objectives, and the principles that support those objectives?

Given an instance of the agile method, does the organization have the ability to support its implementation?

3. *We observe the effect of the instantiated method within the organization.*

   Did the application of the method yield the expected/ intended results?

   Does the organization's supporting environment and/or the adopted practices impede or support producing the expected results?

The problem then is to provide organizations with a systematic approach to assessing their adopted agile method so that they can determine if their method is 'in-sync' with their organizational objectives.

## *1.3 Issues*

In defining an approach to assessing the 'goodness' of agile methods, we have identified the following issues to be addressed:

1. *Determining the assessment perspectives*

   A primary concern in this research has been to ascertain the perspectives from which the 'goodness' of an agile method can be assessed. As discussed previously, we propose to examine (1) a method independent of an organization, (2) an instance of a method within an organization, and (3) the implementation of that instantiated method. This respectively entails (1) assessing the sufficiency of the method with respect to meeting stated objectives, (2) evaluating the ability of an organization to support the implementation of the method, and (3) determining if the method produced the intended results.

2. *Recognizing objectives, principles and practices*

   We recognize the existence of objectives embodied by agile methods, principles used to support the objectives, and practices employed that are reflective of the principles. The agile manifesto [14] provides four focal values that form the core of the agile philosophy and a set of twelve supporting principles. Our task is to derive a set of objectives that are reflective of the agile philosophy and the focal values as stated by the manifesto. Also, from the manifesto, we have to identify a comprehensive set of principles that support the objectives.

3. *Establishing the relationships between the objectives, principles and practices*

This research focuses on determining if an agile method adopted by an organization is consistent with the organizational objectives. Hence, given an agile method with its set of objectives, principles and practices, relationships among them have to be established in order to determine if there are principles that support the objectives and practices that reflect the principles. The relationships between objectives, principles and practices are not explicitly stated in the existing literature. Hence, we have to definitively identify and establish the relationships and provide evidence for the same.

We realize that for an objective and the set of principles associated with that objective, one or more principles may support that objective to a greater extent than the others. Similarly, for each principle and the set of practices that are related to that principle, some practices may be necessary for the achievement of that principle. Recognizing that some relationships between the objectives, principles and practices may be more important than the others, we have to address this disparity in our solution approach.

4. *Identifying the observable characteristics of the people, process, project and product*

Definitive relationships between observable characteristics of the people, process, project, and product, to practices employed by an organization have to be identified. When used, each practice induces an associated set of observable characteristics. We need to identify these characteristics so that we can assess:

- An organization's ability to support the implementation of an agile method.
  - This depends on the people, process and project characteristics.
- The extent to which the method produces the expected results.
  - This depends on the process artifacts and product characteristics

5. *Validating the assessment approach*

Validating the assessment approach primarily involves (1) substantiating the assessment perspectives, and (2) gathering evidence for the existence of the objectives, principles, and practices, the relationships among them, and the observable characteristics. To determine the ability of an organization to support an instantiation of an agile method and evaluate the extent to which that instance produces the expected results, we have to study the method and its implementation within an organization. Hence, to validate the above-mentioned assessment perspectives, we require the application of the assessment approach in an industrial setting.

The solution approach described in the next section outlines our proposed approach to address the above-mentioned issues.

## *1.4 Solution Approach*

We advocate the need for a more comprehensive agile assessment process that assesses the *people, process, project, and product* characteristics of organizations adopting agile methods. In this research, we describe an approach for assessing the 'goodness' of agile methods from three perspectives. More specifically, to assess the collective 'goodness' of a given agile method, we address the following three questions:

- How *adequate* is the method with respect to achieving its objectives?
- How *capable* is an organization in providing the supporting environment to implement its selected method?
- How *effective* is the implementation of the method in achieving its objectives?

In response to the above questions, we propose the **O**bjectives, **P**rinciples and **P**ractices (OPP) framework to facilitate the assessment of the adequacy, capability and effectiveness of agile methods. The framework identifies desirable objectives embraced by the agile philosophy, and definitively links them to principles that support the achievement of those objectives (Figure 1.1). Similarly, the framework also identifies accepted practices and links those practices to the associated principles. We recognize that some linkages between the objectives, principles, and practices are more important than the others. Hence, we assign weights to the linkages.

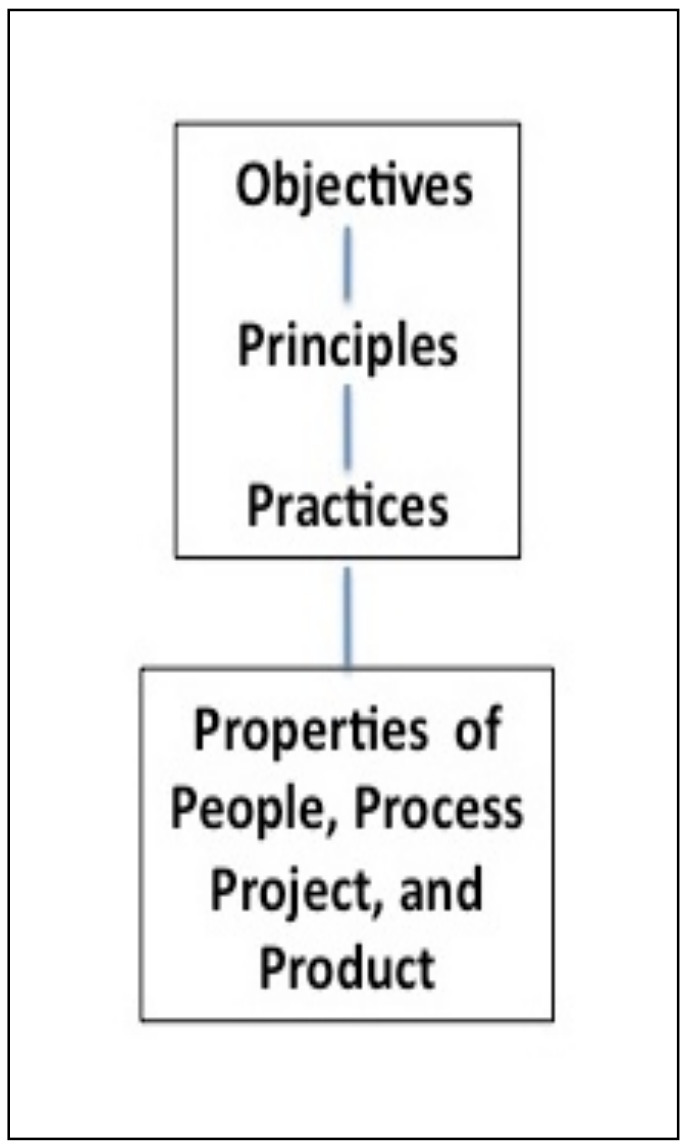

*Figure 1.1. Core structure of the OPP Framework*

The linkages between the objectives and principles, and between the principles and practices, guide the assessment process. We assess the *adequacy* of an agile method by traversing the linkages in a top-down fashion. Given the set of objectives espoused by the agile method, we follow the linkages downward to ensure that the appropriate principles are enunciated, and that the proper practices are expressed. In addition to a top-down examination, the *capability* of an organization to implement its adopted method, and the method's *effectiveness* are assessed using a complementary bottom-up traversal of the linkages. This begins, however, by identifying people, process, project and product characteristics that attest to the use of particular practices. Each practice - property pair is an *indicator*. Then, by following the linkages upward from the practices, we can infer the use of proper principles and the achievement of desired objectives. Figure 1.1 shows the core structure of the OPP Framework. The objectives, principles, practices, properties, and the linkages between them are the foundational pieces of the Framework (Figure 1.1).

We propose to substantiate the components of the OPP Framework by gathering evidence for their existence from literature and collecting feedback from agile practitioners about the Framework itself. We also plan to identify target organizations and assess the adequacy of their customized agile methods and the capability of those organizations to implement the said methods. Assessing the effectiveness of a method would involve a longitudinal study, which is beyond the scope of our proposed validation goals.

## *1.5 Blueprint*

The remainder of this document is organized as follows:

- In Section 2, we review some of the existing agile assessment approaches and the two approaches guiding the OPP Framework.
- Section 3 describes our proposed approach. It provides an overview of the OPP framework and its components.
- Section 4 outlines our approach to assessing 'goodness'.
- In Section 5, we present our proposed substantiation approach.
- Section 6 summarizes our work.

# 2. Background

The need for a structured, systematic and comprehensive approach to assessing the 'goodness' of agile methods forms the basis of our problem statement and solution approach. In this section, we review some of the current agile assessment approaches that have motivated our research. Additionally, to help readers better understand the work presented in this document, we provide background information about the agile philosophy, its values and principles. We also present an overview of the Objectives Principles and Attributes Framework and the Evaluation Environment methodology that guide the construction and design of the OPP Framework (Section 3) and the assessment approach (Section 4).

## *2.1 Agile Overview*

"Software engineering is (1) the application of a systematic, disciplined, quantifiable approach to the development, operation, and maintenance of software; that is, the application of engineering to software, and (2) the study of approaches in (1)" [1]. The Software Development Lifecycle (SDLC) was introduced in order to structure the software development process and ensure that the process complies with the above definition. The software development process systematically begins with identifying and specifying user requirements and then proceeds to the architecting, design, coding, testing and maintenance phases. Traditional software development methodologies such as the Waterfall model [2] and the Spiral model [3] have helped reduce the chaotic state of software development by focusing on:

- Extensive planning to outline the process to be followed, identifying tasks to be completed during product development, and determining product milestones,
- Comprehensive documentation in the form of Software Requirements Specification (SRS), high and low-level design documents, test plans, etc.,
- Gathering and specifying user requirements upfront before proceeding to the architecting phase,
- Anticipating future requirements and designing the architecture of the system to accommodate current and future requirements, and
- Process monitoring and control, risk management, and software quality control and assurance.

However, in the last few years, many development teams have realized that these traditional approaches to software development are not suitable to all [4]. The traditional approaches are found to be inadequate and expensive for development efforts that have to address rapidly changing requirements. Conventional methods attempt to foresee requirements and create the software system architecture upfront in order to accommodate current and future requirements. However, when previously unidentified requirements surface, the current architecture may no longer be valid. The cost of modifying the architecture, and in turn the code, in order to accommodate requirements identified late during the development lifecycle is very high [3].

Creating comprehensive documentation is often a wasted activity in the face of changing requirements. Documents have to be maintained regularly to reflect any changes made to requirements and design. In effect, maintaining comprehensive documentation is expensive. Also, organizations and teams involved in developing small-scale systems find the extensive planning and documentation efforts cumbersome.

The above-mentioned issues prompted the need for new cost-effective, lightweight methods to accommodate rapidly changing requirements in software systems. This realization motivated the development of the "**Agile**" approach, which is best described by the "Manifesto for Agile Software Development" [5] as given below:

"We are uncovering better ways of developing software by doing it and helping others do it. Through this work we have come to value:

**Individuals and interactions** over processes and tools
**Working software** over comprehensive documentation
**Customer collaboration** over contract negotiation
**Responding to change** over following a plan

That is, while there is value in the items on the right, we value the items on the left more"

The core values of the agile manifesto are briefly discussed below:

1. The agile movement focuses on fostering team spirit among the members of the software development team by stressing close relationships and open working environments. Human aspects of the software development process are considered more important than the process itself.

2. The main objective of the software development team is to produce working software at regular intervals. Agile methods advocate the adoption and use of iterative and incremental development approaches. Working software is used to measure progress and minimal documentation is produced.

3. Relationships between the customers and the development team are given preference over strict contracts. Agile methods accentuate delivering software that would provide maximum business value to the customers thereby reducing the risk of not fulfilling the contract. Also, the customer is involved throughout the development process, which fosters better customer-developer relationships.

4. The customers and the development team should be prepared to modify plans in order to accommodate change even late in the development lifecycle.

Agile methods are lightweight processes that focus on *short iterative cycles, direct customer and user involvement, incremental development and accommodating change even late in the development lifecycle*. Also, these methods are value-driven; that is, they focus on maximizing the business value to the stakeholders involved. “Business value is something that delivers profit to the organization paying for the software in the form of an Increase in Revenue, an Avoidance of Costs, or an Improvement in Service (IRACIS) which are discussed below:

- Increase Revenue – Will this functionality increase the revenue generated by the system?
- Avoid costs – Will this functionality improve efficiency and reduce wasteful processes?
- Improve Service – Will this application help us provide timely service and information to others in a better way?” [6].

In the field of software development, there has been shift in focus from conventional software engineering approaches towards agility. The realization that the traditional plan-driven approaches to software engineering are not suitable to all organizations and teams has motivated the creation of the Agile Manifesto. The creation of the manifesto has given rise to many agile methods like eXtreme Programming (XP) [7-9], Scrum [9, 10], Crystal methodologies [9, 11, 12], Feature Driven Development [9, 13], Lean Development [14, 15], etc.

***A generic agile method***

In the following paragraphs, we describe a generic agile method. The information presented here is based on readings from [9, 16-19].

Figure 2.1 shows a generic agile method that follows an iterative and incremental approach. At the start of a project, all the stakeholders meet to decide on its scope. The time required to develop the product is estimated and then divided into multiple release cycles. During each release cycle, a potentially shippable increment of the product is developed. A release cycle is composed of multiple iterations. An iteration is a short software development cycle that lasts two to four weeks as shown in Figure 2.1.

During the initial meeting, often referred to as "***Iteration 0***", the stakeholders identify a set of features. These features are sets of functionality that deliver business value to the customer and are stored on a "***Project Backlog"*** (Figure 2.1). These features are then prioritized and the time required for the development of each is determined. Also, the set of features to be developed during the first release cycle is established.

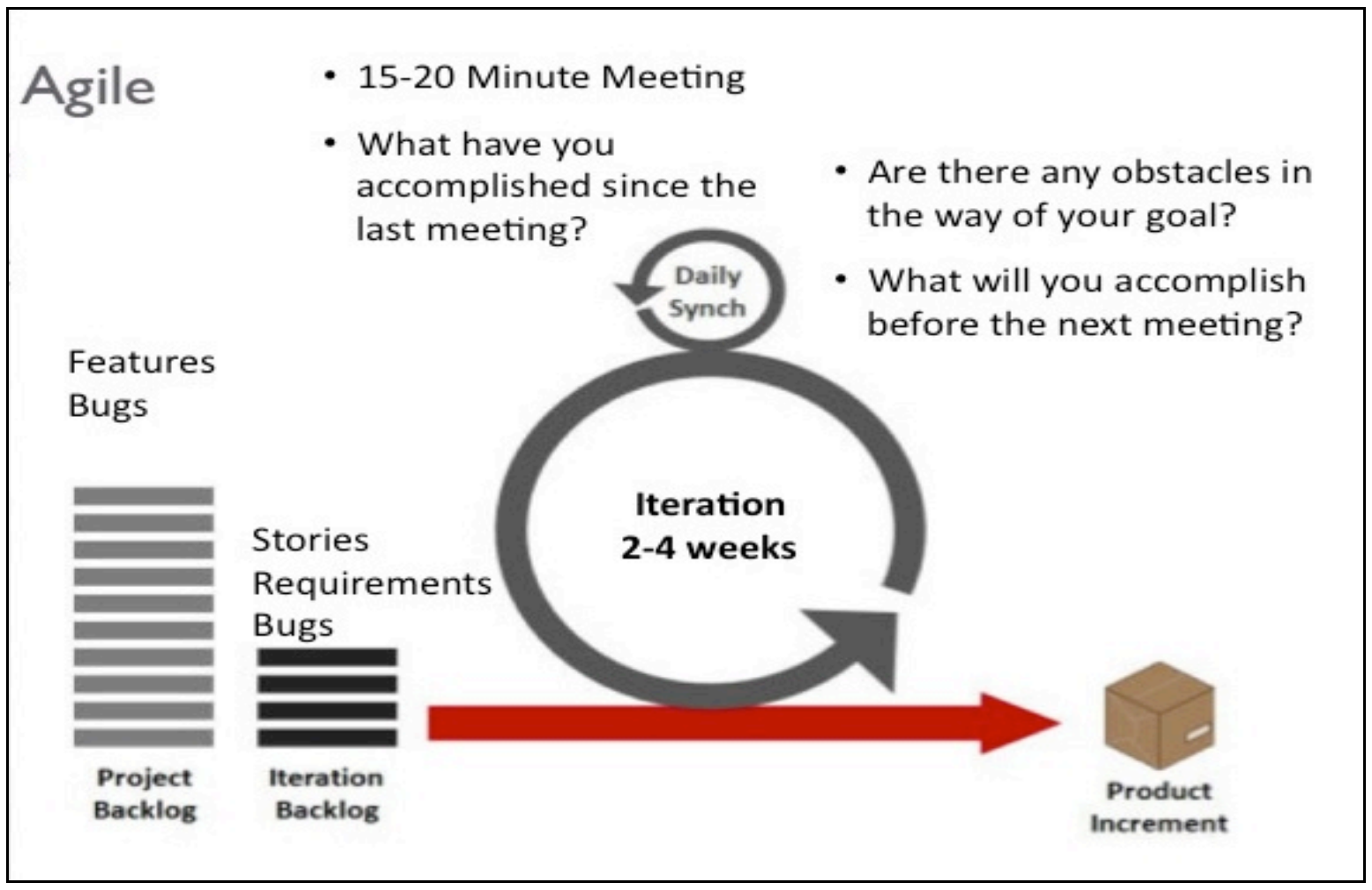

*Figure 2.1 A generic agile method*

Each feature is decomposed into multiple stories. Stories are brief descriptions of user- or customer-valued functionality. The time required to implement each story is determined and the stories are then prioritized. The prioritized stories are stored on the "Iteration Backlog" (see Figure 2.1), which is specific to each iteration. Each story may be further decomposed into tasks. Tasks are essentially checklists for developers, which outline the details required for implementing the stories. During an iteration, the stories are developed and are integrated with the existing code base on a daily basis as indicated by the "Daily Synch" loop in Figure 2.1. Also, as shown in Figure 2.1, the members of the development team meet everyday for 15 to 20 minutes to discuss:

(1) Their accomplishments since the last meeting
(2) The issues (if any) that they faced, and
(3) Their next tasks to be completed.

Agile methods focus on customer communication and collaboration. Therefore, customers or customer representatives are usually present onsite to answer any questions the development team may have and participate in the product development lifecycle. The customers are responsible for creating and recording the acceptance criteria for the stories and features. Developers write tests to meet the acceptance criteria defined by the customers. Also, at the end of each iteration, the customers test the software produced thus far against the acceptance criteria that they outlined previously.

Bugs identified when testing may be stored in the iteration backlog if they can be fixed during the same iteration or moved to the project backlog if they can only be remedied during the next iteration or release cycle.

## *2.2 Agile Assessment*

The agile manifesto states that agile practitioners value "working software over comprehensive documentation" [5]. This value, in conjunction with the agile principle that states "our highest priority is to satisfy the customer through early and continuous delivery of valuable software", primarily guide the choice of agile metrics devised and used [20]. Hence, *most assessment approaches focus on the product*. For example, the number of bugs reported [21], the number of tests written for maximum code coverage [22], the team velocity that indicates the number of story points delivered during each iteration [20], earned business value [11], etc. are some metrics used by agile teams that focus on the product.

While essential to assessing agile methods, the product metrics are by themselves not sufficient to provide a comprehensive assessment. Software Engineering involves *people, process, project,* and *product* (the 4 P's) [23]. Hence, metrics used in the assessment of agile methods should incorporate characteristics of the 4P's. The OPP Framework that we propose in this research addresses this issue. In this sub-section, we review current agile assessment approaches and analyze their merits and demerits.

### *2.2.1. Agile Assessment Checklists*

Agile practitioners are aware that in order to continuously improve an adopted agile method, it is necessary to assess both their process and the final product. As a rule, agile teams are asked 'How agile are you?' In effect, teams are inclined towards ascertaining the extent to which their process is agile. Predominantly, to assess the agility of their process, they use checklists to determine the presence or absence of practices that are considered 'agile'. These checklists, however, ignore the principles reflected by the practices. Also, most of these checklists are tailored to one or more specific agile methods. Some checklists that are commonly used by agile practitioners are (1) the *Nokia Test* [24] *for Scrum*, (2) *How Agile Are You (42-Point Test)* [25], (3) the *Scrum Master Checklist* [26], and (4) the *Do It Yourself (DIY) Project Process Evaluation Kit* [27]. The DIY Project process Evaluation Kit is a generic checklist that can be adapted for use with any agile process. Additionally, these checklists fail to assess the effectiveness of agile methods.

### *2.2.2. Agile Adoption Frameworks*

With the increasing popularity of agile methods, more and more organizations are moving towards agility. However, researchers have realized that many organizations and teams adopt and use agile practices without a proper understanding of the agile philosophy. Moreover, many organizations proclaim themselves to be 'agile' when in fact they are following what is considered "Agilefall" (using agile terms but essentially following the waterfall model) [24]. In order to mitigate this issue and guide organizations in their agile adoption efforts, many agile adoption and process improvement frameworks have been developed.

In his book titled "Balancing Agility and Discipline – A guide for the perplexed", Barry Boehm provides a five step risk-based software development approach that is reflective of the characteristics of both agile and plan-driven development methodologies [28]. This framework is helpful for organizations that require a hybrid approach combining the best aspects of agile and plan-driven software development methods. However, the primary disadvantage of this approach is that the Framework provides an overview of the method to be followed and not actual practices [29]. Organizations require a tangible approach to adopting agile methods. More specifically, they need more guidance about practices that are best suited for their requirements before transitioning to agility. In the next paragraphs, we discuss two frameworks that provide a structured approach to agile adoption and improvement. Both frameworks specify practices that an organization should implement in order to embrace agility.

*Sidky Agile Measurement Index (SAMI)*

SAMI was developed at Virginia Tech by Dr. Ahmed Sidky as an attempt to guide the agile adoption efforts of an organization [29]. The SAMI is based on the idea that the more the number of agile practices adopted by an organization, the greater its agility [29]. The number of agile practices adopted by an organization determines its *agile potential*. Recognizing that counting the number of practices adopted is a simplistic measure, Dr. Sidky has created five agile levels. The agile levels defined in SAMI are very similar to the levels of maturity in CMM and CMMI.

Each agile level has a set of objectives. These objectives reflect the focal values stated in the agile manifesto. Each level is associated with a set of principles stated in the manifesto and practices followed by the agile community [29]. At each level, a set of practices that are reflective of the objectives defined at that level have been identified.

*An organization can choose to achieve any of the specified levels of agility by adopting all the practices at that level and the levels below***.** That is, for an organization to maintain their agility at level 3, all the practices identified for levels 1, 2 and 3 have to be adopted.

SAMI outlines a systematic and structured approach to guide organizations in their agile adoption efforts. The primary disadvantage of the SAMI is that it compromises the flexibility afforded by agile methods. A set of practices for each level is predefined and is "forced" on the organization and thus reducing the flexibility offered by agile methods. Moreover, the established levels may not be reflective of the culture and values of the organizations.

*Agile Adoption and Improvement Model (AAIM)*

Qumer's AAIM [30] is very similar to SAMI in that it provides an agile adoption and improvement framework that specifies varying levels of agility. The intent of AAIM is to measure the degree of agility of an agile method. Agility is defined by five parameters, namely "flexibility, speed, leanness, responsiveness and learning" [31].

Six levels of agility are defined and are further grouped into three agile blocks. Each block and level has an associated set of objectives. At each block, the degree of agility can be measured quantitatively using the agility measurement modeling approach – the 4-DAT (4 Dimensional Analytical Tool) [31, 32].

The 4-DAT has been designed to examine agile methods from four different dimensions: (1) determining the scope of application of the method, (2) agility characterization based on the five parameters mentioned previously - flexibility, speed, leanness, responsiveness and learning, (3) value-based characterization – identifying practices based on the focal values stated in the agile manifesto, and (4) software process characterization – identifying practices covering the SDLC phases, project management, configuration management, and process management [31].

Dimensions 1, 3 and 4 involve a qualitative assessment; dimension 2 involves a quantitative assessment. In dimension 2, the presence or absence of the practices is recorded and the overall score serves as a measure to check the existence of agility in agile methods.

AAIM is modeled along similar lines as the SAMI with respect to the CMMI-like agile levels. As mentioned earlier, the disadvantage is reduced flexibility. The degrees of agility are measured by analyzing the adoption of a set of practices. Also, the predefined levels may not be 'in-sync' with the organizational objectives. Additionally, the AAIM and SAMI do not provide an indication of the effectiveness of the agile approach under consideration.

### *2.2.3. Agility Measurement Approaches*

After transitioning to agility, most organizations are concerned about how 'good' has their agile adoption been - that is, *are they achieving what they set out to by adopting the agile philosophy?* Also, the organizations are interested in identifying problem areas and issues, and take adequate measures to solve them. Retrospective meetings at the end of each iteration or release cycle help an organization or team assess their progress, and 'fine-tune' their agile approach. In addition to retrospection, teams can employ external consultants or tools to help assess their agile process. Most third-party assessment tools/ frameworks are proprietary and information about them is not readily available. Here, we discuss two independent agile assessment frameworks available for free on the web that can be used by agile practitioners to assess the agility of their adopted method.

*Comparative Agility*

Some organizations focus on being more agile than their competition rather than striving to be "perfectly agile" [33]. The Comparative Agility (CA) assessment tool (developed by Kenny Rubin, Mike Cohn, and Dr. Laurie Williams) is used to help organizations assess their agility relative to other organizations or teams that responded to the tool [33]. CA is a survey-based assessment tool. "Any agile practitioner can visit the CA website <http://comparativeagility.com/>, answer the survey questions, and receive a free report that compares their results to the complete industry dataset" [33]. The answers are recorded on a five point Likert scale. Additionally, practitioners could request a customized survey.

At the highest level, CA identifies seven dimensions that form the basis for the agility assessment. These dimensions can be likened to the Key Process Areas (KPAs) identified by the CMM and CMMI. The seven dimensions are: teamwork, requirements, planning, technical practices, quality, culture and knowledge creating [33, 34]. Each dimension is composed of three to six characteristics. Each characteristic has four statements that are assessed by the survey respondents [33]. "Each statement is a practice for which the respondent indicates the truth of the statement relative to their team or organization" [33]. By utilizing a combination of dimensions, characteristics, and statements, a team or an organization, can gauge their agility relative to that of their competitors or themselves at an earlier time.

Through close analysis of the formulated survey questions (available from [34]), it is evident that the statements require an indication of presence or absence of a practice rather than how well is it being used. Respondents base their answers on their observations that are not validated by empirical data. Also, the answers to the survey questions are subjective. Moreover, when comparing the agility of two or more organizations, it is unclear if the tool factors in the differences in their organizational objectives.

*Thoughtworks Agile Assessment*

Thoughtworks is a leading agile development and consulting company. They have developed an agile assessment survey which is available on their website <http://www.agileassessments.com/>. Agile practitioners can complete the survey to get a report on the level of agility within their organization or team, and also identify opportunities for improvement [35]. The survey is composed of twenty questions covering development and management practices [35, 36].

The Thoughtworks Agile Assessment survey questions are intended to gather information about the existence and usage of the practices. This approach, like CA, does not address assessing the effectiveness of agile methods in the survey. However, the creators of the tool recognize and state that the survey is intended to offer *preliminary* insights into the agile method being followed by a team or organization and is not a replacement for a customized assessment approach.

**Summary**

The checklists, the agile process improvement frameworks, and the agile assessment tools discussed in this section primarily focus on the presence or absence of practices (binary values), which is similar to the concept of adequacy that we define in our assessment approach. However, 'adequacy', or any adequacy-like criteria, is not sufficient to provide a comprehensive assessment of an agile method. Adequacy is not indicative of the ability of an organization to implement an agile method or the effectiveness of that method. Hence, we introduce two additional criteria or perspectives for assessing the 'goodness' of agile methods, namely capability and effectiveness. The three perspectives are combined to provide a comprehensive approach for assessing the 'goodness' of agile methods.

## *2.3 Guiding Frameworks*

The OPP Framework and the assessment approach developed in this research are based on the Objectives, Principles and Attributes (OPA) Framework and the Evaluation Environment (EE) Methodology. We present these two approaches in this sub-section. We also preface this discussion with an overview of some of the earlier software measurement approaches.

### *2.3.1. Early Work in Software Measurement*

Software Metrics have been in use since the mid 1960's when the *Lines of Code* metric was adopted as the basis for measuring programming productivity and effort [37]. The rationale for the definition and use of any individual software metric has been either:

"(a) the desire to assess or predict effort/cost of development processes, or
(b) the desire to assess or predict quality of software products" [37].

Hence, most Software Measurement approaches adopted by organizations utilize metrics to assess or predict the quality of software products. Metrics to assess quality involve the assessment of software quality objectives such as maintainability, reliability, testability, correctness, adaptability, etc.

The primary goal of Software Engineering is to produce a quality product [38]. The abstract nature of software has contributed to the difficulties in assessing its quality. Consider the following examples discussed in [39] and [40] respectively:

1. If the goal is to assess maintainability, we may conjecture that as the number of unconditional branches in a program increases, the more difficult it would be to maintain the program.

2. Schedule slippage adversely impacts the quality of the software product. However, one might incorrectly surmise that if a project is on schedule, the system being built is a quality product.

In both cases mentioned above, the measures lack the definitive linkage that directly relates the measure to software quality. That is, the measures discussed here are marginally related to the goal. Hence, predicting the software quality using such measures that are "somewhat" related to the goal is difficult. Recognizing the need for defined measures to be directly related to software quality, researchers developed metrics and measurement programs that utilized attributes of the process to assess the product. McCall's Factor/ Criteria/ Metric [39], Goal Question Metric (GQM) [41], the Objectives Principles and Attributes (OPA) Framework [40, 42, 43], and the Evaluation Environment (EE) Methodology[44-47] are a few assessment approaches that have been developed to address the above- mentioned issue.

Consider the "Goal Question Metric" (GQM) method. "Goal-Question-Metric (GQM) is a paradigm for the systematic definition, establishment, and exploitation of measurement programs supporting the quantitative evaluation of software processes and products." [48].

The GQM method defines a measurement model consisting of three levels [41]:

- Conceptual level **(GOAL) -** A goal defined for an ***object*** *(that part of the product or process being observed)* based on entities like

  a) ***purpose*** *( the motivations for studying the object)*,
  b) ***issue*** *(the characteristic of the object under study)* and
  c) ***viewpoint*** *(people interested in studying the object)*

- Operational level **(QUESTION) -** A set of questions formulated that must be answered to determine if the goals have been reached.

- Quantitative level **(METRIC) -** A set of metrics derived from the questions to collect data to answer the same quantitatively. The collected data could be objective or subjective.

The GQM approach to goal-oriented measurement consists of a *top-down decomposition or refinement* of the goals into questions and then into metrics and a *bottom-up analysis and interpretation* of the collected data [49]. The data collected by the use of metrics are used to answer the formulated questions, which in turn determine if the stated goals have been met. Here, there is an established relationship between the goal, the questions and the metrics. Hence, we can definitively predict and/or assess quality characteristics using metrics that have been formulated to address specific questions regarding those characteristics.

#### *2.3.2. Objectives Principles Attributes (OPA) Framework*

The OPP Framework that we propose for the assessment of the ‘goodness’ of agile methods is based on the OPA Framework. The OPA Framework (developed at Virginia Tech) is a structured and systematic approach that serves as a basis for the evaluation of Software Engineering Methodologies [42]. The OPA Framework identifies:

- Project-level software engineering ***objectives*** like maintainability, reliability, correctness, etc. that are commonly recognized by various methodologies
- ***Principles*** that are reflective of the software development process such as information hiding, structured programming, etc, and
- ***Attributes*** that result from the utilization of the said principles such as readability, traceability, etc. [40].

“Achievement of the objectives comes through the application of principles supported by a methodology. Employment of the said principles result in software products possessing attributes considered desirable and beneficial” [42]. Also, definitive linkages between these components are defined. In addition to the above-mentioned components, the framework identifies properties that are observable characteristics of the product. These properties are indicative of the extent to which the product possesses the desirable attributes. Each attribute is linked to one or more properties. Each attribute, property pair forms an indicator. The indicators form the basis for defining metrics. The values obtained by the usage of these metrics attest to the existence of desirable software engineering attributes in the product [42]. These values are propagated and aggregated at the principles level to present an assessment of the development process, and then further aggregated at the objectives level to suggest the extent to which the objectives are being achieved.

A Software Engineering methodology is assessed from two different perspectives – adequacy and effectiveness. Adequacy denotes the extent to which a methodology can support the achievement of stated project-level goals and objectives [42]. Assessing the adequacy involves a top-down examination of the linkages between the objectives, principles, and attributes to attest to the existence of those Framework components. The evaluation of the adequacy of a methodology is intended to reveal its deficiencies [42]. While a top-down traversal of the linkages is adopted for assessing the adequacy of a methodology, the effectiveness of that methodology is determined by a top-down and bottom-up examination through the linkages [42]. "Effectiveness of a methodology can be defined as the degree to which that methodology produces the desired results" [42]. Analyzing the metric values imply the degree to which particular properties are observed. In turn, this information can be used to identify a set of attributes present. If these attributes are different from the set of attributes identified during the adequacy assessment, it can be implied that the users have failed to adhere to the principles supported by the methodology [42].

### *2.3.3. Evaluation Environment Methodology*

Evaluation of software systems is becoming increasingly difficult because the systems themselves are large and complex. Moreover, the process involves the measurement and evaluation of hundreds of qualitative and quantitative elements, mandates evaluations by Subject Matter Experts (SMEs), and requires the integration of disparate measurements and evaluations [44]. Hence, "planning and managing such evaluation efforts require a unifying methodology and should not be performed in an ad-hoc manner" [44]. In order to address the issues in evaluating large and complex systems, the Evaluation Environment (EE) methodology has been developed [44, 46, 47] at Virginia Tech by Dr. Osman Balci. The EE methodology has been designed to conduct evaluations on different types of projects classified under various domains. In order to accommodate geographically distributed teams, a Web-based client/server software system has also been developed. The Web-based system can be accessed from <https://www.orcacomputer.com/ee/>. The EE Methodology mandates evaluations by SMEs. Some of the main characteristics of the EE Methodology that are adopted by the OPP Framework are presented below [45]:

1. ***Indicators*** - "Qualitative concepts" [44] such as maintainability, reliability, correctness, etc. cannot be measured directly. The EE methodology proposes to use a hierarchy of indicators to measure quality attributes. Each "qualitative concept" that cannot be measured directly is a *root indicator* [44]. Hence, to measure each "qualitative concept", we decompose the root indicator into multiple levels of child indicators such that the indicators at the lowest level can be assessed directly. The indicators at the lowest level are called the *leaf indicators*. *Branch indicators* are those that have a parent indicator and at least one child indicator. We assess only the leaf indicators directly.

2. ***Decomposition*** – As mentioned in the previous paragraph, we decompose a root indicator into the branch and leaf indicators. The multiple levels of indicators form a hierarchy that is an acyclic graph. Also, "each leaf indicator is manageable in complexity and is directly assessable by way of testing, direct measurement, analysis, or examination" [45]. Only the leaf indicators are measured and evaluated.

3. ***Integration*** - When there are multiple leaf indicators for a root indicator, the SMEs critically weigh the leaf indicators by performing a pair-wise comparison using the Analytical Hierarchy Process (AHP). More often than not, the leaf indicators are assigned equal weights. Leaf indicators are evaluated by a variety of techniques such as testing, direct measurement, analysis, and examination. All the evaluations are then integrated under a structured framework. Leaf indicator evaluation scores are integrated in a bottom-up fashion based on the criticality weightings of the branch indicators. The integration results in an overall evaluation score for the root indicator [44, 45]. This final score would help us evaluate the root indicator.

### *2.3.4. Relationship of the OPA Framework and EE Methodology to the OPP Framework*

The OPA Framework and the EE Methodology guide the structure of the OPP Framework and the assessment approach developed in this research. Both approaches present hierarchical assessment frameworks where qualitative concepts such maintainability, reliability, correctness, etc. are evaluated by using defined measures that are directly related to those concepts. We adopt the decomposition strategy touted by both frameworks to assess the extent to which an agile method achieves its stated objectives. The OPA Framework provides the concept of objectives, principles, practices, and linkages and the assessment perspectives of adequacy and effectiveness. Similar to the EE Methodology, our assessment of capability and effectiveness involve the evaluation of people, process, project, and product properties. We develop our indicator hierarchy and the bottom-up assessment approach utilizing the indicator scores based on the EE Methodology.

# 3. Evolving the Assessment Framework

Our research is motivated by the lack of a comprehensive approach to assessing agile methods. We assess the collective 'goodness' of an agile method adopted by an organization based on (1) its *adequacy*, (2) the *capability* of the organization to provide the supporting environment to implement the method, and (3) the method's *effectiveness*. We define adequacy, capability and effectiveness as below (definitions adapted for current context from [1-3]:

- *Adequacy* - Sufficiency of the method with respect to meeting stated objectives
- *Capability* – Ability of the organization to provide the supporting environment conducive to the implementation of the method - which is dependent on its people, process and project characteristics.
- *Effectiveness* – Producing the intended or expected results - which is dependent on process and product characteristics

We have designed the **O**bjectives **P**rinciples and **P**ractices (**OPP**) Framework to assess the 'goodness' of an agile method from the three perspectives described above. This chapter provides an overview of the OPP Framework and its formulated components.

## *3.1. The Framework*

Figure 3.1 provides an overview of the OPP framework. The OPP framework identifies (1) *objectives* of the agile philosophy, (2) *principles* that support the objectives, (3) *practices* that are reflective of the principles, (4) the *linkages* between the objectives, principles and practices, and (5) *indicators* for each practice to assess the effectiveness of the practice and the extent to which the organization supports its implementation.

The culture of an organization, its values and desired characteristics of the systems that it builds, determine the objectives, principles and practices that it adopts. Our assessment of an agile method is carried out with respect to satisfying those stated objectives. Figure 3.1 illustrates the relationships among the objectives, principles, practices and properties. This relationship is central to our assessment process and is common to the assessment of adequacy, capability, and effectiveness.

To assess the adequacy of an agile method, we traverse the linkages in a top-down manner from objectives to principles and from principles to practices. The existence of principles and practices supporting the desired objectives is indicative of the adequacy of the method under consideration.

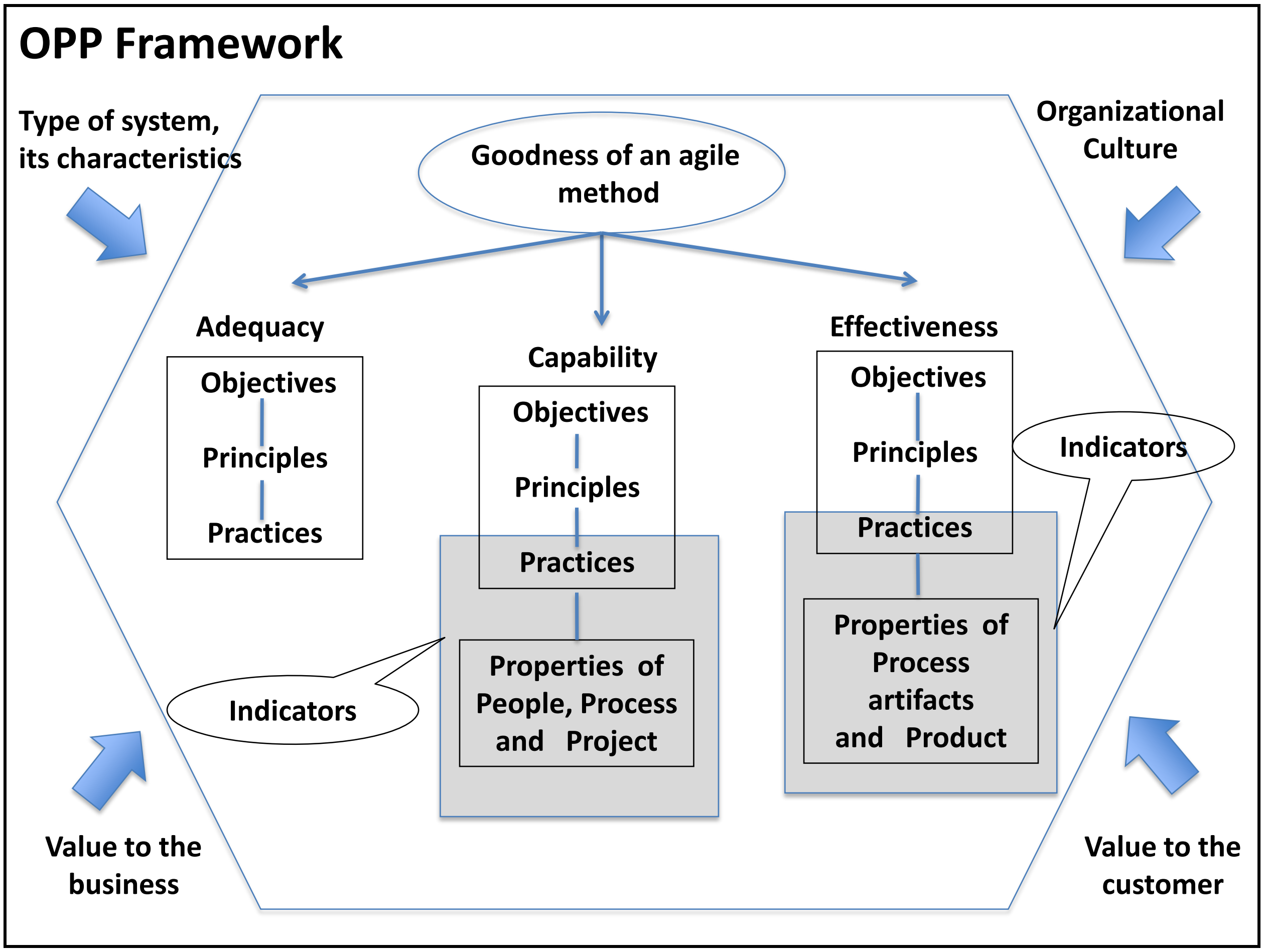

*Figure 3.1 The Objectives, Principles and Practices (OPP) Framework*

Additionally, the OPP framework identifies *people, process, project, and product* properties of the practices adopted. These observable characteristics are associated with the practices. *Each (practice, property) pair is called an indicator*. These indicators are essential to the assessment of capability and effectiveness. Figure 3.1 also shows the indicators below the practices for assessing capability and effectiveness.

*People*, *process* and *project* properties of a practice imply the presence (or absence) of characteristics needed in the supporting environment, i.e., capability. Similarly, other *process artifacts* and *product* properties of a practice are representative of the expected results (effectiveness). Both sets of properties are used to effect a bottom-up assessment of capability and effectiveness by traversing the linkages from the appropriate properties to practices, practices to principles, and principles to objectives. We discuss the components of the OPP framework in the following sub-section.

## *3.2. Formulated Components of the OPP Framework*

At the heart of the OPP framework are the objectives, principles, practices and the linkages that tie them together. Indicators are identified and are required for the assessment of capability and effectiveness. The tasks necessary to sufficiently define the components of the OPP framework are as follows:

### *1. Deriving the objectives and identifying the supporting principles and the practices*

We recognize that each agile method embodies a set of objectives, which are supported by a set of underlying principles. Also, there are practices adopted that are reflective of those identified principles. However, these objectives are not explicitly stated. The agile manifesto provides four focal values and twelve principles that define the agile philosophy. Our work involves deriving objectives reflective of the agile philosophy from the focal values, identifying principles that support the defined objectives, and identifying practices that reflect the principles.

As illustrated in Figure 3.2, we have identified an initial set of objectives based on the agile philosophy. A supporting aggregated set of principles has also been identified from sources including, but not limited to, the agile manifesto [4], books [5-7], research papers [8, 9], experience reports, white papers [10] and discussions with industry experts. Likewise, we have identified a list of practices embraced by the agile community.

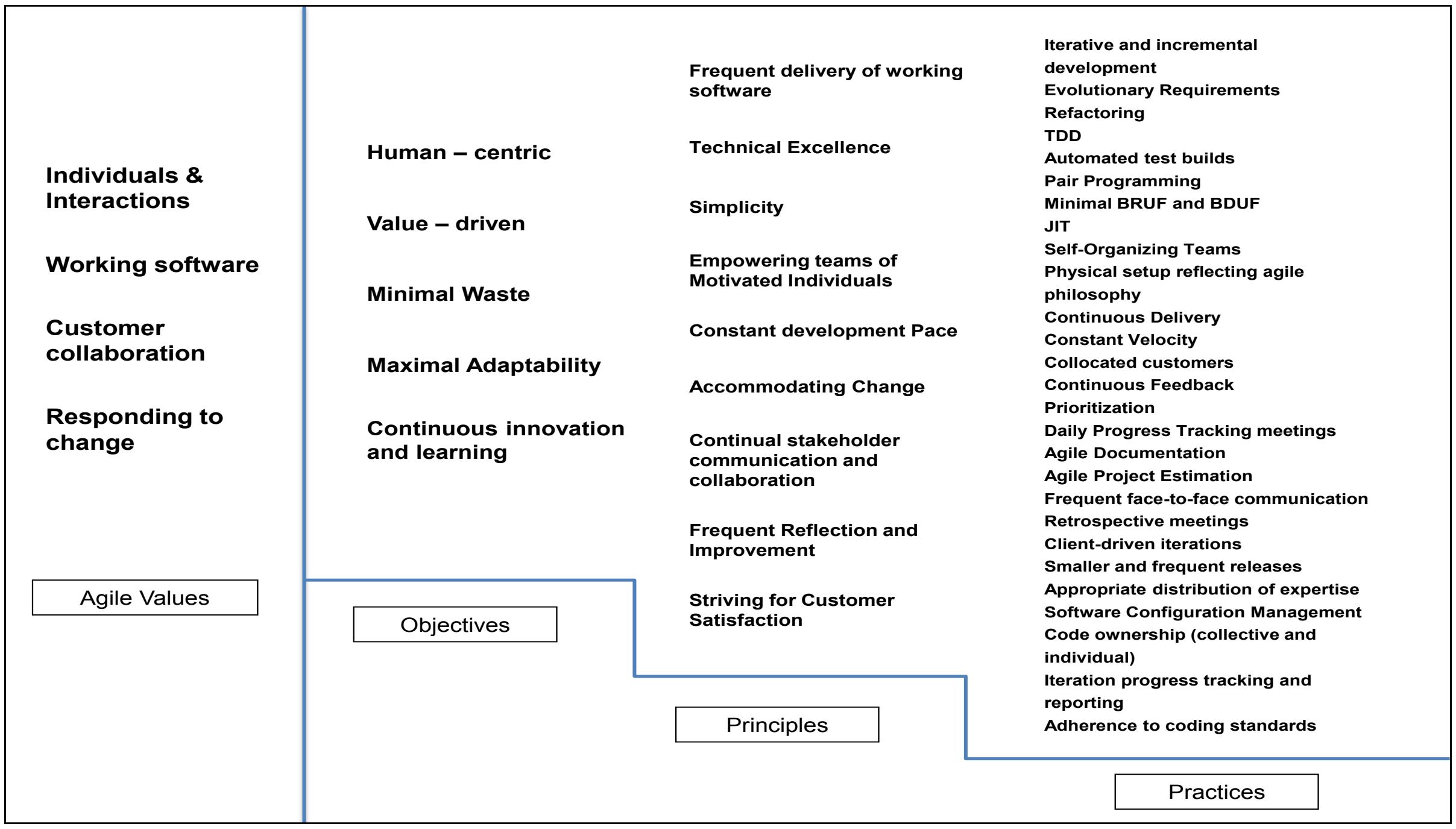

*Figure 3.2. Objectives, Principles, and Practices*

The objectives, principles and practices are the foundational pieces used to assess the adequacy, capability and effectiveness of an agile method. See Appendix 3.1 for a list of working definitions for the objectives and principles used in this research.

***2. Establishing definitive linkages between the identified objectives, principles and practices***

Linkages between the objectives, principles and practices have to be defined in order to assess the adequacy, capability and effectiveness. These linkages represent definitive relationships between the foundational pieces. For example, let us assume that an organization lists *maximal adaptability* (see Figure 3.2) as one of its objectives.

Our working definition for maximal adaptability is "flexibility, ability to accommodate change and having the freedom to choose practices with respect to the people, process, project, and product" (see Appendix A). Hence, one underlying principle that supports maximal adaptability is *accommodating change.* Subsequently, as shown in Figure 3.3, there exists a linkage between the objective 'Maximal Adaptability' and the principle 'Accommodating Change'. We then have a set of practices such as *evolutionary requirements* [11], *iterative and incremental development* [12], *on-site or co-located customer* [13], *continuous feedback* [14], and *minimal Big Requirements Up Front (BRUF)* [11] that help realize the principle of 'Accommodating Change' (also shown in Figure 3.3).

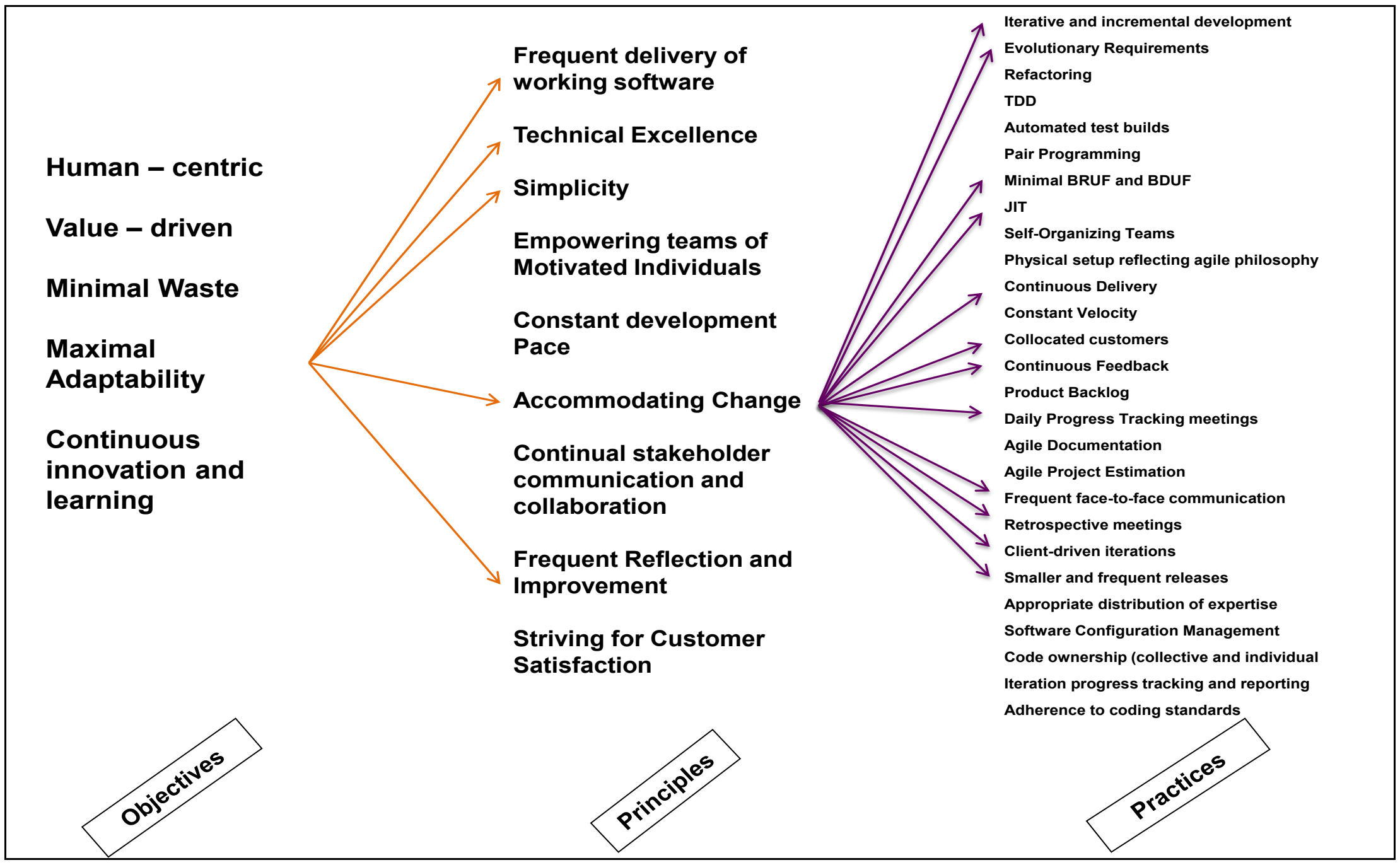

*Figure 3.3. Example linkages in the OPP Framework*

Following the same systematic process described previously, we have identified a candidate set of linkages between all objectives, principles and practices. The complete set of the identified linkages is shown in Figure 3.4. We have used learning, experience reports, white papers, books, etc., to identify all and confirm many of those relationships. Appendix B provides an alternate representation of the candidate set of identified linkages.

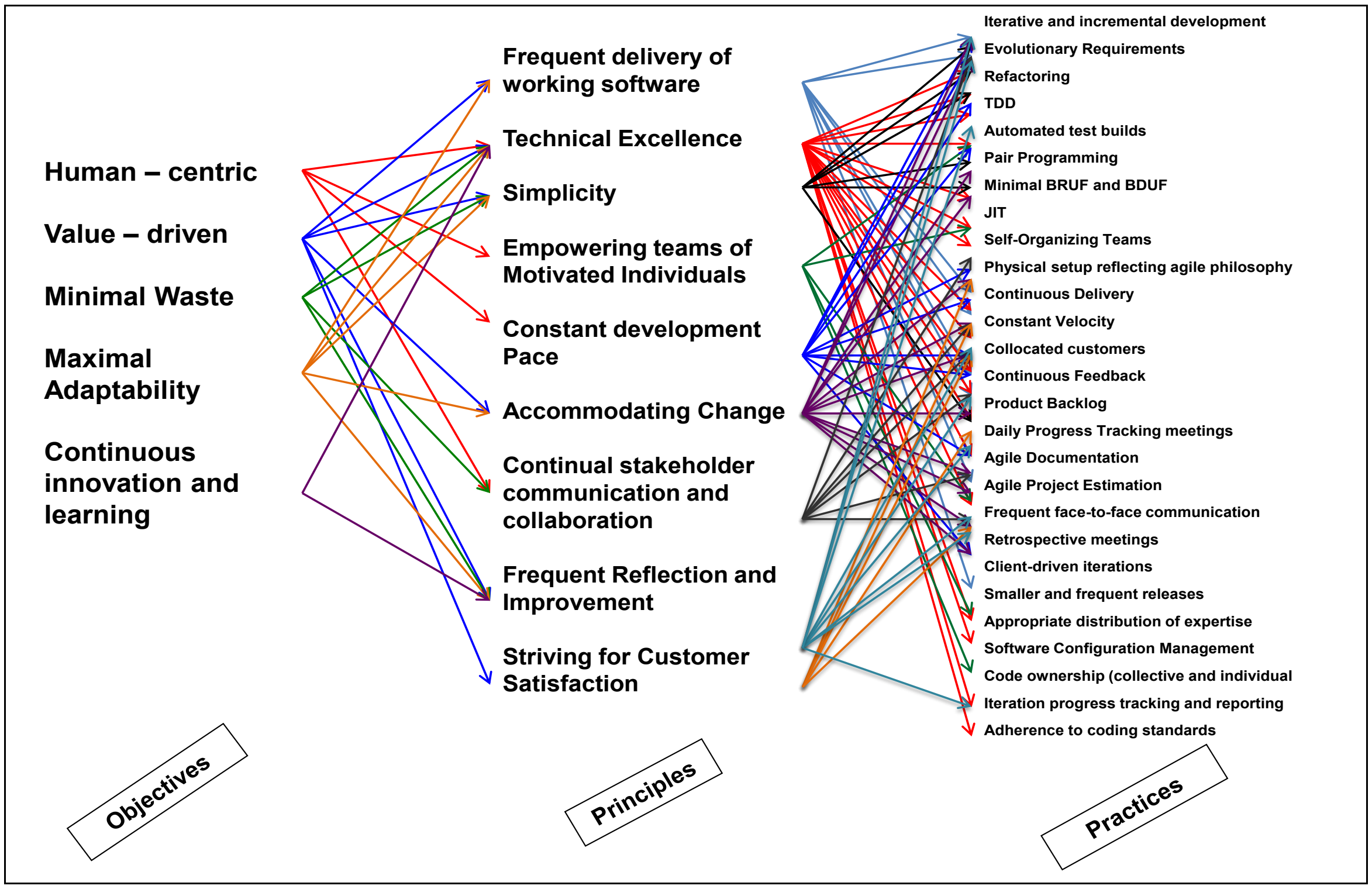

*Figure 3.4. Candidate set of linkages in the OPP Framework*

Although not shown in Figures 3.2, 3.3, and 3.4, the OPP framework supports an additional level of linkages between every practice and a set of properties that are germane to the implementation of the practice are established to assess capability and effectiveness. Those indicators and linkages are discussed next.

### 3. *Identifying indicators*

As discussed previously, to assess (i) the capability of an organization to support the implementation of an agile method and (ii) the effectiveness of the method itself, we propose a bottom-up traversal of the established linkages. At the lowest level of the OPP framework we have the practices and hence we begin our assessment of capability and effectiveness from the practices. However, *how do we determine if the organization has the supporting environment to effectively implement a practice and/ or if the practice has produced the intended results?* The answer lies in the identification of *observable properties* of the practices. These properties are characteristics of the *people, process, project, and product*, and are specific to each practice. A practice and property pair forms an *indicator*. The first step in defining the indicators is to identify the observable properties of the people, process, project, and product associated with each practice.

As discussed below, indicators are required to assess

i. the capability of the organization, and

ii. the effectiveness of the method

*Capability*

We define the capability of an organization as its ability to provide the supporting environment conducive to the implementation of an agile method. In assessing the capability of an organization, we are concerned with the characteristics of its *internal environment*. The internal environment of an organization is primarily composed of its resources and competencies. More specifically, in an organization, the characteristics of its people, the process (including the physical environment) that it adopts, and its projects are reflective of the characteristics of its internal environment. Hence, we use observable properties of the people, process and project in our assessment of capability. For example, the presence of open physical environments in an organization is *indicative* of the organization's capability to foster face-to-face stakeholder communication and collaboration at any given time.

We identify observable properties of the people, process and project for every practice by asking the following questions:

1. What special skills or knowledge do the people involved in the project need to successfully adopt and implement the practice?
2. What characteristics of the process and/or the environment extend support for the implementation of the practice?
3. Are there any project specific characteristics that support or impinge on the effective realization of the practice?

Consider the agile practice of *pair programming* [15]. This is one of the more popular practices where programmers develop code in pairs. At any given time, only one programmer can write code. He or she is called the *driver.* The other programmer, called the *observer* or *navigator*, provides assistance to the driver and reviews the code continuously. The programmers switch roles at regular intervals. Asking the questions mentioned above with respect to pair programming provides us with the following characteristics:

1. People and their skill sets:

   - Willingness of the people to work as pairs and to share their knowledge with their partners.

2. Processes and Physical Environment:

   - Availability of only one computer terminal for two programmers working as a pair.

*Effectiveness*

An agile method is judged effective if it produces "*working software*" that is a *quality product*, *on time*, *within budget*, and *of value to the customers*. Assessing the effectiveness of the agile method involves the identification of properties of process artifacts and the product, which focus on the system produced and its value to the stakeholders. For example, agile teams strive to maintain constant velocity (the empirical observation of the work done by a team during an iteration [16]). Velocity is reflective of the efficiency and effectiveness of the team. Indicators of constant velocity can be obtained by studying Velocity, Burn-Up and Burn-Down charts.

We ask the following questions to determine the observable properties of process artifacts and the product:

1. Are there any process artifacts generated when implementing the practice? If so, what are the criteria for the verification and validation of these artifacts?
2. What are the criteria to determine the successful implementation of the practice? What properties are observed in the final product that is a direct result of the successful implementation of the practice under consideration?

Let us revisit the agile practice of pair programming discussed previously. Asking the questions mentioned above to determine the properties of the product and process artifacts with respect to pair programming provides us with the following:

1. Product:

   - Continuous review of code

   - Knowledge sharing among programmers working in pairs

The output is the code itself developed in a modular fashion.

We have identified an initial (albeit, incomplete) set of observable properties that we present in Appendix C.

#### *4. Defining metrics for the indicators*

Once the observable properties are identified, our next step is to define assessment metrics for each indicator. The measures obtained will be indicative of the capability of the organization to support the implementation and the effectiveness of the practices. For example, the number of bugs reported is a metric associated with continuous code review. As mentioned above, continuous code review is a product property associated with pair programming. Hence, the number of bugs reported can be used to assess the effectiveness of pair programming. If the measure is less than the average number of bugs reported, then we can conclude that pair programming is effective with respect to continuous code review.

Our bottom-up approach to assessing the capability and effectiveness follows the process outlined by the EE methodology [17, 18]. Given a practice, we first identify multiple observable properties that attest to its implementation. Then, for each (practice, property) pair, i.e. for each indicator, we evolve a metric. To determine the extent to which that practice is being used, we aggregate the values of those metrics associated with that (practice, property) pair. After determining the extent to which the practices of a method are effectively employed, we now have to ascertain the extent to which the principles touted by the agile method are indeed reflected by the practices. This is achieved by a further aggregation of the degree to which the set of practices associated with each principle are used. Similarly, a further bottom-up analysis from the principles to the objectives is carried out in order to assess the extent to which the stated objectives are achieved.

## *3.3 Proposed Work*

The OPP Framework provides the foundation for our assessment of the 'goodness' of agile methods. The objectives, principles, practices, linkages, and indicators are the crucial components of the framework. Now that we have identified the components and established the linkages, we can assess the adequacy of agile methods. However, in order to assess the capability and effectiveness, the appropriate indicators have to be identified and the metrics defined. In this research, with respect to the OPP Framework, we propose the following tasks:

- The set of objectives, principles and practices that we have identified is a working set. We need to ***revisit*** them periodically to ensure that they are still necessary and sufficient.
- Currently, we have identified a preliminary set of linkages between the objectives and principles and principles and practices. These linkages definitively identify the relationships between the foundational pieces of the OPP Framework. However, we have defined these linkages through reasoned conjecture. Hence, we need to ***establish*** their completeness and validity, and propose to do so by gathering evidence from existing literature and confirmation by practitioners in the field.

- We describe indicators as a (practice, property) pair. Our first step towards defining the indicators is to ***identify*** the set of observable properties for each practice. The list of observable properties that we currently have is incomplete and we are currently involved in identifying the complete set.
- For each indicator that we define, there is an associated set of metrics to assess the usage of a practice. We still need to ***define*** the metrics for each indicator in order to facilitate capability and effectiveness assessment.

# 4. Our Proposed Approach to Assessing the 'Goodness' of Agile Methods

Given the OPP Framework described in Section 3, we assess the 'goodness' of an agile method by evaluating its adequacy, the capability of the organization to provide the supporting environment to implement the method, and the effectiveness of that method. The Framework provides us with the foundational pieces, namely the objectives, principles, practices, linkages, and indicators needed to guide our approach to assessing 'goodness'.

## *4.1. Assessing Adequacy*

Adequacy is defined by the sufficiency of an agile method with respect to meeting its stated objectives. It is *independent* of an organization. More specifically, we can assess the adequacy of standalone agile methods such as eXtreme Programming (XP) [1-3], Lean [4], Crystal [5], Feature Driven Development (FDD) [1, 2, 6], etc. with respect to the agile values and principles each espouses. That is, given an objective of a method, are the necessary principles also present that are prescribed by the Framework? Then, for each principle enunciated by the Framework, are the recommended practices touted by the agile method? If necessary principles and practices are missing, then adequacy is suspect.

We begin assessing the adequacy of an agile method by analyzing and identifying its objectives, principles, and practices. We determine the linkages between its foundational pieces. The adequacy assessment process is a top-down approach that does not involve indicators. From the linkages in the OPP Framework, we ascertain the expected number of linkages between each objective and its associated set of principles, and each principle and the practices that are reflective of that principle. We then determine the *linkage coverage* for each objective and principle, which is the ratio between the actual number of linkages found and the expected number of linkages. We map the linkage coverage to a Likert scale to determine the adequacy of the method under consideration with respect to its objectives and principles. Based on the linkage coverage and the Likert scale values, we assess the adequacy of a method with respect to its objectives.

### *Weighted Linkages*

When formulating our approach to assessing adequacy, we realized that with respect to the linkages between the objectives and principles, some principles are reflective of an objective to a greater extent than the others. Consider the objective 'Maximal Adaptability'. From Table B.2 (see Appendix B), we know that 'Maximal Adaptability' is supported by the following principles:

- Frequent Delivery of Working Software
- Technical Excellence
- Simplicity
- Accommodating Change, and
- Frequent Reflection and Learning

Here, 'Accommodating Change' is more reflective of 'Maximal Adaptability' than the others. Hence, it carries more weight than the other principles that are associated with that objective. For our assessment, we assign such linkages a *weight that is twice that of the linkages between the other principles associated with that objective*. Here, 'Accommodating Change' is assigned a weight of **2**, and the other principles a weight of **1** each with respect to "Maximal Adaptability". Table 4.1 below shows the linkages between the objectives and principles. The entries in bold represent linkages that carry a weight twice that of the other linkages.

*Table 4.1. Weighted Linkages*

| ***Objectives/Principles*** | ***Freq. Delivery*** | ***Tech. Excellence*** | ***Simplicity*** | ***Emp. Teams of motivated individuals*** | ***Const. Development Pace*** | ***Acc. Change*** | ***Continual stkhlder comm. and coll.*** | ***Freq. refln and learning*** | ***Striving for customer satisfaction*** |
|---|---|---|---|---|---|---|---|---|---|
| Human-centric | | X | | **X** | X | | **X** | | |
| Value-driven | **X** | X | X | | | X | | X | **X** |
| Minimal waste | | X | **X** | | | | X | X | |
| Maximal Adaptability | X | X | X | | | **X** | | X | |
| Continuous Innovation and Learning | | X | | | | | | **X** | |

These weighted linkages play a major role in the assessment of the extent to which the touted objectives are supported by a method.

Similarly, some practices reflect the 'spirit' of a principle to a greater extent than the others. More specifically, there is a set of necessary and sufficient practices that are associated with each principle. Current work is underway to assign weights to the linkages between principles and practices.

A summary of the steps involved in assessing the adequacy of an agile method is given below:

1. We examine the agile method under consideration, and identify its objectives, principles, and practices.

2. From the OPP Framework, we determine the linkages that exist between the objectives, principles and practices stated by the method.

3. From the OPP Framework, we obtain the expected number of linkages between the identified objectives, principles, and practices (weighted for the linkages between the objectives and principles). More specifically, if the method under consideration touts 'Continuous Learning and Innovation' as one of its objectives, from the OPP Framework, we know that number of linkages associated with that objective is *two* and the weighted count is *three* (Table 4.1). Hence, for the above-mentioned objective, the expected number of weighted linkages is three.

4. We establish the linkage coverage as a ratio of the number of linkages present (from the method under consideration) and the number of linkages expected (from the OPP Framework). Currently, we use weighted linkage coverage only for the linkages between objectives and principles. We simply count the number of linkages between the principles and practices.

5. After determining the linkage coverage for each objective and principle, we map the values to a Likert scale. Presently, we use the criteria given below in Table 4.2 to map the percentage value of the linkage coverage to the Likert scale values. The criteria and the Likert scale values used here are subjective and are not yet final.

*Table 4.2. Linkage coverage criteria and Likert scale values*

| Linkage Coverage Criteria | Likert Scale Values |
|---|---|
| 81 - 100 % | Strongly |
| 61 – 80% | Very well |
| 41 – 60% | Sufficiently |
| 21 – 40% | Somewhat |
| 0 – 20% | Very little |

Currently, for each objective and principle, we provide a Likert scale value based on the criteria given in Table 4.2. These values are indicative of the extent to which the method being assessed is adequate with respect to each objective and principle that it states. We plan to outline an approach for representing the adequacy assessment results.

We have assessed the adequacy of three agile methods – FDD, XP, and Method A. A detailed account of the assessments is provided in Section 4.3.

## *4.2. Assessing Capability and Effectiveness*

Unlike adequacy, we cannot assess the capability and effectiveness of a standalone agile method that is independent of an organization, because both capability and effectiveness require an organizational perspective. Assessing the capability of an organization to support the implementation of a method and the effectiveness of the method itself requires both a top-down (objectives → principles → practices) and bottom-up (properties → practices → principles → objectives) traversal of the linkages. *The difference between the assessment of capability and effectiveness lies primarily in the type of indicators used* - people, process and project indicators for capability, and process artifacts and product indicators for effectiveness.

It is required to ensure the existence of a practice before proceeding to assess the extent to which it is implemented in an organization. Also, it is necessary to confirm that the practice is supported by the organization before determining the degree to which it is being used. Hence, within an organization, we assess the adequacy of the adopted agile method, followed by the capability of the organization to implement that method, and finally the method's effectiveness.

Our bottom-up approach to assessing the capability and effectiveness follows the process outlined by the EE methodology [7, 8]. Given a practice, we first identify multiple observable properties (people, process, project and product characteristics) that attest to its implementation. Then, for each (practice, property) pair, i.e. for each indicator, we evolve a metric. To determine the extent to which that practice is being used, we aggregate the values of those metrics associated with all the practice - property pairs specific to that practice. The aggregated values are indicative of

- the extent to which the implementation of the practice is supported by the organization, and
- the effectiveness of the practice itself.

After determining the extent to which the practices of a method are effectively employed, we assess the extent to which the principles touted by the agile method are indeed reflected by the practices. This is achieved by a further aggregation of the degree to which the set of practices associated with each principle are used. Similarly, a further bottom-up analysis from the principles to the objectives is carried out in order to assess the extent to which the stated objectives are achieved.

We recognize that the metrics to be defined for each indicator would yield values that maybe subjective, objective, numerical, binary, range values, etc. Hence, we need to map the different types of values obtained onto a *uniform scale of measurement* to perform the aggregation. However, we have not yet defined that necessary mapping approach. Current research is underway to identify an appropriate uniform scale of measurement. The EE Methodology suggests one viable approach.

## *4.3 Preliminary Work and Findings*

In this research, we propose to assess agile methods from three perspectives – adequacy, capability and effectiveness. The adequacy of an agile method is assessed independent of any organizational objectives. Unlike adequacy, the capability of an organization to support the implementation of an agile method and the effectiveness of the method itself, are assessed from an organizational perspective. Our preliminary work focuses on adequacy. More specifically, we have examined the adequacy for two standalone methods – XP [1-3] and FDD [6], and for Method A that is an instantiation of XP in an organization. XP is an agile method that best reflects the agile philosophy. Method A is an XP variant implemented in an organization. FDD is intended for medium to larger scale systems development and hence touts a blend of agile values and conventional software engineering principles.

The following sub-sections summarize the assessment results that we have obtained by applying our adequacy assessment process to FDD, XP and Method A. For each of the three methods, we identified their touted objectives, principles and practices, and established the linkages between them. We then determined the linkage coverage and mapped the values to the Likert scale described in Section 4.1.

### *4.3.1. FDD Adequacy Assessment*

Feature Driven Development (FDD) is an agile method developed by Jeff De Luca and Peter Coad. It is an iterative and incremental approach that is more structured than the other agile methods and is designed for scalability. Hence, it is commonly used for developing medium to larger scale systems. In the words of Peter Coad, "FDD has just enough process to ensure scalability and repeatability and encourage creativity and innovation all along the way" [1].

FDD does not address the process of requirements engineering. It focuses only on the design and coding phases of development and does not span the complete development process. Therefore, other Requirements Engineering methods have to be used in conjunction with the FDD approach to developing software.

*Adequacy Assessment*

As outlined in Section 4.1, we begin the assessment approach by analyzing the method under consideration. We identify the objectives touted by the method, the supporting principles, and the adopted practices, and identify the linkages between them. As discussed previously, we assign weights to the linkages between the objectives and principles. For each stated objective, we compute the weighted linkage coverage that is the ratio between the observed weights for the linkages and the expected weights.

*Observations about FDD*

The success of FDD, like other agile methods, is largely due to recognizing software development as a human endeavor. Hence, 'good' people become the primary asset to the organization adopting FDD. The creators of FDD are of the opinion that the people are more important than the process.

FDD is diametrically opposite to the other agile methods in its practice of upfront modeling. The other methods, XP for example, declare that the architecture of a system should emerge from an iterative process over several increments. FDD on the other hand places emphasis on "getting it right the first time"[1, 6]. Hence, during the initial modeling phase, a lightweight architecture is created and the class skeletons are written.

The major disadvantage of the upfront modeling and design efforts is that these practices reduce the ability to accommodate change later in the development cycle. The architecture of the system is 'frozen' at the beginning of the development effort, and accommodating major changes would be expensive. However, *the fact that FDD supports upfront modeling and design is the primary reason for its suitability for developing medium and larger scale systems*.

The FDD approach to developing software is not "lean" [9]. More specifically, it proposes to build more than what is needed.

FDD does not accommodate client driven iterations. Proponents of FDD believe in the development team driving the iterations as opposed to the other agile methods where the clients stipulate the order in which the features are developed.

Based on our analysis and observations, we have identified a set of objectives, principles, and practices that we conjecture are touted by FDD. Table 4.3 provides a list of objectives and principles stated by the method. Also, the linkage coverage and the weighted linkage coverage values are included in Table 4.3.

*Table 4.3. Linkage coverage for FDD*

| Objectives | Linkage Coverage | Weighted Coverage |
|---|---|---|
| Human-centric | 3/ 4 | 4/ 6 |
| Value-driven | 5/ 6 | 7/ 8 |
| Maximal Adaptability | 3/ 5 | 3/ 6 |
| Continuous Innovation and Learning | 2/ 2 | 3/ 3 |

| **Principles** | **Linkage Coverage** |
|---|---|
| Frequent Delivery of Working Software | 8/ 8 |
| Technical Excellence | 12/ 18 |
| Empowering teams of motivated individuals | 4/ 5 |
| Constant Development Pace | 7/ 10 |
| Frequent Reflection and Improvement | 6/ 9 |
| Striving for Customer Satisfaction | 3/ 5 |

Based on the linkage coverage values provided in Table 4.3, we map these values for FDD to a Likert scale as described previously in Section 4.1. Tables 4.4 and 4.5 provide the mappings of the linkage coverage values to the Likert scale for objectives and principles respectively.

*Table 4.4. Adequacy assessment of FDD with respect to its touted bbjectives (using weighted linkage count)*

| | Strongly | Very well | Sufficiently | Somewhat | Very little |
|---|---|---|---|---|---|
| **Human-centric** | | X | | | |
| **Value-driven** | X | | | | |
| **Maximal Adaptability** | | | X | | |
| **Continuous Innovation and Learning** | X | | | | |

From Table 4.4, we know that among the objectives touted by FDD, 'Value-driven' and 'Continuous Innovation and Learning' are supported to a greater extent than the others. As mentioned previously, FDD does not focus on 'being lean'. That is, FDD does not state the objective of 'Minimal waste' and its associated principle of 'Simplicity'. FDD is intended for medium to larger scale systems development and is more structured than the other agile methods. More specifically, its construction and design minimizes the ability to accommodate change. Hence, FDD supports 'Maximal Adaptability' to a lesser extent (Table 4.4).

*Table 4.5. Adequacy Assessment of FDD with respect to its stated principles (using linkage count)*

| Principle | Strongly | Very well | Sufficiently | Somewhat | Very little |
|---|---|---|---|---|---|
| **Frequent delivery of working software** | X | | | | |
| **Technical Excellence** | | X | | | |
| **Empowering teams of Motivated Individuals** | | X | | | |
| **Constant development pace** | | X | | | |
| **Frequent Reflection and Improvement** | | X | | | |
| **Striving for Customer Satisfaction** | | | X | | |

Table 4.5 provides an overview of the extent to which FDD supports its stated principles. FDD is an iterative and incremental software development approach that focuses primarily on 'Frequent Delivery of Working Software' and touts necessary practices that are reflective of the said principle (Table 4.5). As mentioned previously, customers are not involved throughout the development lifecycle and the iterations are not client-driven. That is, 'Striving for Customer Satisfaction' is supported to a lesser extent as shown in Table 4.5. Teams maintain a constant development pace. At regular intervals, the stakeholders meet to reflect and 'fine-tune' their process.

*Summary*

As discussed previously, FDD is more suited for medium to larger scale systems. It is more structured than the other agile methods. By design, the method does not state certain objectives such as 'Minimal Waste' and 'Maximal Adaptability'. Also, certain classic agile practices such as 'Client-driven Iterations', 'Refactoring', 'Minimal Big Requirements Up Front and Big Design Up Front', and 'Test Driven Development' are not supported. Hence, some of the touted objectives and principles are achieved to a lesser extent. Also, from our adequacy assessment results for FDD, XP and Method A, we surmise that FDD is the least adequate with respect to its stated objectives than the other two methods.

#### *4.3.2 XP Adequacy Assessment*

eXtreme Programming (XP) is designed for small to medium teams which have to develop software quickly in the face of rapidly changing requirements. XP is based on four values: *communication, simplicity, feedback and courage* [2, 3] and "It works by bringing the whole team together in the presence of simple practices, with enough feedback to enable the team to see where they are and to tune the practices to their unique situation" [2]. The customers are actively involved in the development process. XP came into existence with software development practices found effective during the previous decades.

*Observations about XP*

The design of the system is usually kept simple.

Developers work in pairs and implement features using Test Driven Development (TDD).

The customers write acceptance criteria against which the system is tested at the end of each iteration and release cycle.

Each iteration lasts 2 - 4 weeks.

XP is a set of practices that can be adopted by agile practitioners who may choose to use some or all of the practices or may customize the method to suit their needs.

XP is designed for small teams. However, certain XP practices can be used in larger scale systems development as well (for example, pair programming).

*Adequacy Assessment*

Table 4.6 provides a summary of the linkage coverage and weighted coverage values for the objectives and principles supported by XP.

*Table 4.6. Linkage coverage for XP*

| Objectives | Linkage Coverage | Weighted Coverage |
|---|---|---|
| Human-centric | 4/ 4 | 6/ 6 |
| Value-driven | 6/ 6 | 8/ 8 |
| Minimal Waste | 4/4 | 5/5 |
| Maximal Adaptability | 5/ 5 | 6/ 6 |
| Continuous Innovation and Learning | 2/ 2 | 3/ 3 |

| Principles | Linkage Coverage |
|---|---|
| Frequent Delivery of Working Software | 8/ 8 |
| Technical Excellence | 15/ 18 |
| Simplicity | 6/ 6 |
| Empowering teams of motivated individuals | 4/ 5 |
| Constant Development Pace | 10/ 10 |
| Accommodating Change | 11/ 11 |
| Continual stakeholder communication and collaboration | 6/7 |
| Frequent Reflection and Improvement | 8/ 9 |
| Striving for Customer Satisfaction | 5/ 5 |

Mapping the linkages coverage values for XP from Table 4.6 to the Likert scale used in this research (see Section 4.1), we obtain the nominal values shown in Tables 4.7 and 4.8.

*Table 4.7. Adequacy assessment of XP with respect to its touted objectives (using weighted linkage count)*

| | | | | |
|---|---|---|---|---|
| **Human-centric** | | | | |
| **X** | | | | |
| Strongly | Very well | Sufficiently | Somewhat | Very little |
| **Value-driven** | | | | |
| **X** | | | | |
| Strongly | Very well | Sufficiently | Somewhat | Very little |
| **Minimal Waste** | | | | |
| X | | | | |
| Strongly | Very well | Sufficiently | Somewhat | Very little |
| **Maximal Adaptability** | | | | |
| X | | | | |
| Strongly | Very well | Sufficiently | Somewhat | Very little |
| **Continuous Innovation and Learning** | | | | |
| X | | | | |
| Strongly | Very well | Sufficiently | Somewhat | Very little |

XP is an agile method that best reflects the objectives of the agile philosophy. All of its stated objectives are maximally supported by associated principles (Table 4.7).

*Table 4.8. Adequcy assessment of FDD with respect to its stated principles (using linkage count)*

| | | | | |
|---|---|---|---|---|
| **Frequent delivery of working software** | | | | |
| X | | | | |
| Strongly | Very well | Sufficiently | Somewhat | Very little |
| **Technical Excellence** | | | | |
| | X | | | |
| Strongly | Very well | Sufficiently | Somewhat | Very little |
| **Simplicity** | | | | |
| X | | | | |
| Strongly | Very well | Sufficiently | Somewhat | Very little |
| **Empowering teams of Motivated Individuals** | | | | |
| | X | | | |
| Strongly | Very well | Sufficiently | Somewhat | Very little |
| **Constant development pace** | | | | |
| X | | | | |
| Strongly | Very well | Sufficiently | Somewhat | Very little |
| **Accommodating Change** | | | | |
| **X** | | | | |
| Strongly | Very well | Sufficiently | Somewhat | Very little |
| **Continual stakeholder communication and collaboration** | | | | |
| | **X** | | | |
| Strongly | Very well | Sufficiently | Somewhat | Very little |
| **Frequent Reflection and Improvement** | | | | |
| | **X** | | | |
| Strongly | Very well | Sufficiently | Somewhat | Very little |
| **Striving for Customer Satisfaction** | | | | |
| **X** | | | | |
| Strongly | Very well | Sufficiently | Somewhat | Very little |

By definition, XP values ‘Simplicity’ and ‘Accomodating change’. From Table 4.8, we know that XP is maximally adequate with respect to the said principles. The method also values effective communication and continuous feedback. Iterations are client-driven and teams strive to maximize customer satisfaction (Table 4.8).

*Summary*

XP best reflects the agile philosophy. It is designed for small teams. From Tables 4.7 and 4.8, we know that XP supports its stated objectives and principles to a large extent. The method is designed to accommodate change even late in the development lifecycle. Based on our assessment, we conjecture that XP is the most adequate with respect to its touted objectives when compared to Method A and FDD.

### *4.3.3 Method A Adequacy Assessment*

Organization X is a Fortune 1000 company that provides software development, agile coaching and IT consulting services. The method under consideration in this sub-section, Method A is a proprietary agile method of Organization X. As mentioned previously, Method A is an instantiation of XP. It embodies almost all of the values and principles stated by XP. However, the culture of the organization precludes the adoption of certain touted principles and practices. A former employee of Organization X who had adopted Method A for over two years provided us with an overview of the method. Given the OPP Framework as a baseline, the following are some of our observations.

*Observations about Method A*

With respect to the objectives, Method A primarily focuses on being human-centric and delivering value to all the stakeholders involved. Method A touts continuous learning and innovation as one of its objectives but, in reality, the teams do not support this objective.

Great emphasis is placed on delivering working software at regular intervals, maintaining a constant development pace, continual stakeholder communication and collaboration, and striving to satisfy the customers. Simplicity and frequent reflection and learning are principles that have the least support.

XP mandates that customers be co-located in order to achieve face-to-face communication and collaboration at any given time. However, in reality, co-location though preferred, is not always feasible. Hence, to overcome this issue, Method A stipulates that customers be present during the first and last days of an iteration. Communication with the customer is achieved via other means such as video conferencing during other times.

Method A mandates the adoption of Continuous Integration which is "*a software development practice where members of a team integrate their work frequently, usually each person integrates at least daily - leading to multiple integrations per day"* [10]. Also*, "each integration is verified by an automated build (including test) to detect integration errors as quickly as possible. Many teams find that this approach leads to significantly reduced integration problems and allows a team to develop cohesive software more rapidly."* [10].

Customer/ User Acceptance Testing (CAT/ UAT) is performed on the last day of an iteration. The customers / users test the system built thus far and determine if the acceptance criteria have been met.

Bug fixes and enhancements to the system are classified as new tasks and are performed immediately.

Adopting Test Driven Development (TDD) is optional. Hence, the team members may or may not write the tests before writing code.

The teams are not self-organizing. The management plays a key role in creating the teams for various projects.

*Adequacy Assessment*

Table 4.9 provides a summary of the linkage coverage and weighted coverage values for the objectives and principles touted by Method A.

*Table 4.9. Linkage coverage for Method A*

| Objectives | Linkage Coverage | Weighted Coverage |
|---|---|---|
| Human-centric | 4/ 4 | 6/ 6 |
| Value-driven | 5/ 6 | 7/ 8 |
| Minimal Waste | 4/4 | 5/5 |
| Maximal Adaptability | 4/ 5 | 5/ 6 |

| Principles | Linkage Count |
|---|---|
| Frequent Delivery of Working Software | 8/ 8 |
| Technical Excellence | 15/ 18 |
| Simplicity | 5/ 6 |
| Empowering teams of motivated individuals | 4/ 5 |
| Constant Development Pace | 10/ 10 |
| Accommodating Change | 11/ 11 |
| Continual stakeholder communication and collaboration | 6/7 |
| Striving for Customer Satisfaction | 4/ 5 |

We map the linkage coverage values presented in Table 4.9 to the Likert scale that we have been using for our assessment purposes (Section 4.1). The resulting Likert scale values are provides in Tables 4.10 and 4.11.

*Table 4.10. Adequacy assessment of Method A with respect to its touted objectives (using weighted linkage count)*

| | Strongly | Very well | Sufficiently | Somewhat | Very little |
|---|---|---|---|---|---|
| **Human-centric** | **X** | | | | |
| **Value-driven** | **X** | | | | |
| **Minimal Waste** | **X** | | | | |
| **Maximal Adaptability** | **X** | | | | |

As shown in Table 4.10, Method A, like XP, maximally supports all of its stated objectives. By design, Method A does not state ‘Continuous Innovation and Learning’. As discussed previously, within the organization, the definition of the adopted method differs from the method in its original form. Hence, Method A, which is an instantiation of XP, does not state certain objectives and principles defined in the baseline method.

*Table 4.11. Adequacy assessment of Method A with respect to its stated principles (using linkage count)*

| | Strongly | Very well | Sufficiently | Somewhat | Very little |
|---|---|---|---|---|---|
| **Frequent delivery of working software** | **X** | | | | |
| **Technical Excellence** | | **X** | | | |
| **Simplicity** | | **X** | | | |

| | Strongly | Very well | Sufficiently | Somewhat | Very little |
|---|---|---|---|---|---|
| **Empowering teams of Motivated Individuals** | | X | | | |
| **Constant development pace** | X | | | | |
| **Accommodating Change** | X | | | | |
| **Continual stakeholder communication and collaboration** | | X | | | |
| **Striving for Customer Satisfaction** | | X | | | |

Method A does not support 'Frequent Reflection and Learning'. The other stated principles are supported maximally (Table 4.11). Method A focuses on delivering working software frequently, maintaining a constant development pace, facilitating continual stakeholder communication and collaboration, and striving to satisfy the customers as shown in Table 4.11.

*Summary*

Method A is an instance of XP. Its application within an organization precludes the adoption and use of certain principles and practices. Comparing the adequacy assessment results of XP (Tables 4.7, 4.8, and 4.9) and Method A (Tables 4.9, 4.10, and 4.11), we can observe the differences in their stated objectives and principles. Method A does not support the objective 'Continuous Innovation and Learning' and its associated principle 'Frequent Reflection and Improvement'. The other touted objectives and principles are maximally achieved. In assessing the adequacy of XP, FDD, and Method A, we recognize that Method A with respect to its stated objectives is (1) less adequate than XP and (2) more adequate than FDD. That is, it falls in between XP and FDD with respect to supporting its stated objectives and principles.

## *4.4. Proposed Work*

Currently, we have designed the OPP Framework to guide our approach to assessing adequacy, capability, and effectiveness. The tasks identified next have to be completed with respect to the assessment approach.

**1. Assessing Adequacy:**

We follow a top-down approach in traversing the linkages between the objectives, principles, and practices to assess adequacy. To evaluate the adequacy of any given agile method, we still need to

- ***define*** the final criteria for mapping the linkage coverage values to a Likert scale.
- ***assign*** weights to linkages between principles and practices
- ***revisit*** the adequacy assessment results after assigning weights to the linkages between principles and practices and calculate the values again.

We also plan to ***assess*** the adequacy of at least two more agile methods. Also, we propose to rank the methods based on their adequacy. Such a ranking scheme would help us assess the extent to which each method supports its stated objectives.

**2. Assessing Capability and Effectiveness:**

Assessing the capability of an organization to support the implementation of an agile method, and the effectiveness of that method, involve a top-down and bottom-up traversal of the linkages established between the objectives, principles, and practices. Presently, we have a skeletal structure of our approach to assessing capability and effectiveness. To define and implement our approach, the following tasks have to be completed.

- We need to ***identify*** observable characteristics of the people, process, project and product associated with each practice. The primary difference in the two assessment perspectives is the types of indicators used. We use people, process, and project indicators to assess capability and process artifacts and project indicators to assess effectiveness. As discussed previously, each indicator is a (practice, property) pair. See Appendix C for a preliminary list of observable characteristics.

- We need to ***define*** metrics for the evaluation of each indicator. The values produced by the metrics are then utilized in computing the extent to which each practice is used by an organization.

- The metrics may produce values that are subjective, objective, numerical, binary, etc. These values have to be aggregated in order to ascertain the degree of usage of a practice. Hence, we need an approach to map the varying types of values produced by the metrics to a uniform scale.

- Currently, we do not possess a definitive approach to representing the results obtained when assessing the adequacy of an agile method, the capability of an organization in supporting the implementation of that method, and the method's effectiveness. We need to describe a method that would enable us to represent the results obtained.

Though we have outlined the approach to assessing the capability and effectiveness, we first need to identify properties of the people, process, project, and product associated with the practices, and the appropriate metrics, in order to effect a bottom-up traversal of the linkages to assess agile methods from the said perspectives.

## *Selected Bibliography*

# 5. Substantiating the OPP Framework

The OPP Framework guides the assessment process. Our goal is to substantiate both the components of the OPP Framework and our process for assessing adequacy, capability and effectiveness. The research outlined in this section is work in progress.

## *5.1. Substantiating the components of the OPP Framework*

The objectives, principles, practices, and the linkages between them, form the core of the OPP Framework. The Framework also identifies people, process, project and product characteristics that are necessary to assess the capability of the organization and the effectiveness of the method under consideration. We outline the following 2-step approach to substantiating the components of the OPP Framework:

*Step 1:* Gathering evidence from literature and other sources to

- *Substantiate the linkages between objectives, principles and practices*

  We are currently involved in gathering evidence from research papers, experience reports, white papers and books to validate the existence of the proposed linkages. Our efforts in research, learning and interactions with members of the agile community provide us with the necessary evidence for the existence of the identified linkages. We have currently substantiated approximately 50% of the identified linkages.
- *Substantiate the weights assigned to the linkages between the objectives, principles, and practices*

  We realize that for each objective and its set of associated principles, some principles may support that objective to a greater extent than the others. Hence, we have assigned weights to the linkages between the objectives and principles. We are currently in the process of apportioning weights to the linkages between the principles and practices. We plan to request subject matter experts to evaluate our weights distribution scheme and provide feedback.

- *Confirm the indicators*

  We are in the process of identifying all the (practice, property) pairs that form the indicators. We then propose to substantiate the indicators by gathering evidence from literature and discussions with members of the agile community.

*Step 2:* Validate the components of the OPP Framework by obtaining feedback from agile practitioners through presentations and using survey instruments.

We propose to gather feedback from members of the agile community about the OPP Framework and its utility. We have identified four target organizations around the Blacksburg, VA area where we intend to present our work and request the practitioners to complete a survey that would provide us with feedback. The survey questions would focus on the utility of the OPP Framework, its components and the assessment process.

We have spoken to representatives from the target organizations and they have been receptive to participating in our substantiation efforts.

### *5.2. Substantiating the assessment process*

We address the assessment of agile methods from three perspectives – adequacy, capability and effectiveness. We realize that in order to effectively substantiate our assessment approach, the OPP Framework has to be applied within an organization. We envision a two-step process for substantiating and validating the assessment approach.

*Step 1:* Using the OPP Framework, we first propose to assess the adequacy of multiple agile methodologies endorsed by the agile community (XP, Scrum, Lean, etc.).

Currently, we have assessed the adequacy of XP, FDD and Method A. Method A is an instantiation of XP within an organization. *Our preliminary assessment results indicate that XP is more adequate than FDD and Method A with respect to its stated objectives. Also, FDD is the least adequate with respect to its touted objectives.*

We propose to assess the adequacy of two other standalone methods namely Lean, and Crystal.

*Step 2*: Secondly, we intend to apply the OPP Framework within multiple organizations to assess

1. the *adequacy* of its agile method and
2. the *capability* of the organization to provide the supporting environment to implement that agile method

Gathering feedback from practitioners is an effort to substantiate our approach and assessing the 'goodness' of agile methods adopted by organizations is a validation process.

In addition to gathering feedback about our assessment approach, we propose to assess the adequacy of agile methods adopted by one or more of our target organizations and capability of those organizations to support the implementation of the said methods. We intend to interview project managers and request that they give us a walkthrough of their process. We also plan to observe one or more teams in action to understand their approach to developing software. This would provide us with key insights about the process followed and the supporting environment, and thus facilitating the assessment of adequacy of the method and capability of the organization. After assessing the adequacy of the agile methods and the capability of the target organizations, we propose to gather feedback about the assessments to attest to the validity of our results.

While validating our approach to assessing the effectiveness of an agile method is necessary, it requires a longitudinal study that falls beyond the scope of our immediate validation goals.

# 6. Conclusion

Our research is motivated by the need for a comprehensive approach to assess the ‘goodness’ of agile methods. We assess ‘goodness’ of an agile method based on (a) its adequacy, (b) the capability of an organization to provide the supporting environment for implementing the method, and (c) the effectiveness of that method. To guide our assessment, we propose the OPP Framework. The Framework identifies objectives, principles, practices and properties, and linkages between them to support the assessment process. Currently, we have

- Designed the OPP Framework,
- Identified objectives, principles, and practices, and determined the linkages between them,
- Defined the approach to assessing adequacy, capability, and effectiveness,
- Outlined the necessary substantiation efforts, and
- Assessed the adequacy of three agile methods – FDD, XP and Method A.

We have determined the linkages between the objectives, principles, and practices based on reasoned conjecture. In order to establish their completeness and validity, we propose to substantiate them by gathering evidence from literature and feedback from agile practitioners. Current work is underway to confirm the defined linkages. To assess capability and effectiveness, we still need to identify people, process, project, and product properties associated with the practices. For each (practice, property) pair, or indicator, we propose to define metrics for assessing the extent to which the said practices are being used. The values obtained will be propagated in a bottom-up fashion to determine the capability of an organization to support the method’s stated objectives, and the effectiveness of the method itself. Our proposed substantiation approach includes a study of one or more organizations to assess the ‘goodness’ of their agile methods, at least from an adequacy and capability perspective. We recognize that the Framework would evolve based on future research considerations.

# Appendix A

Here we provide the working definitions (Table A.1) for the objectives and principles identified as a part of the OPP Framework. These definitions have stemmed from our understanding and in-depth analysis of the agile philosophy and methods currently used.

*Table A.1. Working definitions for the objectives and principles*

**Objectives**

***Human – centric***

People are more important than processes, practices and tools

***Value-driven***

Maximize value to all the stakeholders. Value may take the form of increased revenue, improved customer satisfaction, etc.

***Minimal Waste***

Keep things simple. Build only what is necessary.

***Maximal Adaptability***

Flexibility, ability to accommodate change and having the freedom to choose practices with respect to the people, process, project, and product.

***Continuous innovation and learning***

Frequently reflect on the process and learn from past mistakes to innovate and improve.

**Principles**

***Frequent delivery of working software***

Deliver working software frequently; iteration length – 2 to 4 weeks.

***Technical Excellence***

Excellence of the people, process and practices – selecting the right people, right process and right practices to build working software of value to the customer. Technical excellence is also conveyed through the supporting environment.

***Simplicity***

Of the process and product – keep the process simple and produce a product containing only what is necessary.

***Empowering teams of Motivated Individuals***

Build teams of motivated individuals and empower them; the decision-making process is taken to the lowest level.

***Constant development Pace***

Build software at a constant pace; the amount of work achieved during each iteration should be constant.

***Accommodating Change***

Accommodate change with minimal impact.

***Continual stakeholder communication and collaboration***

The stakeholders interact and work together at regular intervals to discuss the system being built.

***Frequent Reflection and Improvement***

Revisit the process regularly; retrospective meetings help in learning more about the existing process and support improvement.

***Striving for Customer Satisfaction***

The main objective is to satisfy and provide maximum value to the customer.

# Appendix B

Tables B.1 and B.2 provide an alternate representation of the identified linkages between objectives and principles, and principles and practices respectively. An "X" mark or a number within square brackets in a table cell denote the existence of a linkage. Our approach has been to gather evidence from literature and other sources to establish the validity of some of the linkages. The numbers represented within square brackets denote the reference number of the source from which the evidence can be found. The references can be found at the end of this section. The cells with the "X" mark represent linkages that have not been substantiated yet.

*Table B.1. Identified linkages between objectives and principles*

| ***Objectives/Principles*** | ***Freq. Delivery*** | ***Tech. Excellence*** | ***Simplicity*** | ***Emp. Teams of motivated individuals*** | ***Const. Development Pace*** | ***Acc. Change*** | ***Continual stkhlder comm. and coll.*** | ***Freq. refln and learning*** | ***Striving for cust. satisfaction*** |
|---|---|---|---|---|---|---|---|---|---|
| Human-centric | | [7][18] | | [1][7] | X | | [7][18] | | |
| Value-driven | X | [11] | X | | | X | | X | X |
| Minimal waste | | X | X | | | | X | X | |
| Maximal Adaptability | X | X | X | | | X | | X | |
| Continuous Innovation and Learning | | X | | | | | | X | |

*Table B.2. Identified linkages between principles and practices*

| ***Practices/Principles*** | ***Freq. Delivery*** | ***Tech. Excellence*** | ***Simplicity*** | ***Emp. Teams of motivated individuals*** | ***Const. Development Pace*** | ***Acc. Change*** | ***Continual stkhlder comm. and coll.*** | ***Freq. refln and learning*** | ***Striving for cust. satisfaction*** |
|---|---|---|---|---|---|---|---|---|---|
| Iterative and Incremental Devt. | X | | | | [8] | [5] | | X | |
| Evolutionary Reqts. | X | | X | | X | [16] | [15] | X | |
| Refactoring | | [24] | X | | | | | | |
| TDD | | [28] | [28] | | [15] | | | | |
| Automated test builds | | X | | | | | | X | |
| Pair Programming | | [10][14] | | [7][10] | [10][14] | | | | |
| Minimal BRUF, and BDUF | | | X | | | X | | | |
| JIT | | X | X | | | X | | | |
| Self-Organizing teams | | [12][25][27] | | X | | | | | |
| Working environment reflective of agile philosophy | | X | | | | | X | | |
| Continuous Delivery | X | | | | X | X | | | [27] |
| Constant Velocity | X | X | | | X | | | | |
| Collocated customers | | [6] | | | | [25] | [25] | [25] | X |
| Continuous Feedback | [6] | [5] | | | [5][12] | [3][12] | [8] | [6] | X |
| Prioritization | X | X | | | X | | | | |
| Daily Progress Tracking Meetings | | X | | | | X | X | [22] | |
| Agile Documentation | | X | X | | | | | | X |
| Agile Project Estimation | X | | | | X | | | | |
| Frequent face-to-face communication | | [6] | | X | | X | X | X | |
| Retrospective meetings | | | | | | X | | [23] | |
| Client-driven iterations | | | | | | | X | | X |
| Smaller and frequent releases | [1] | | | | [8][12][17] | [5][12] | | | |
| Appropriate composition of expertise | | [19][20] | | [19][20] | | | | | |
| ***Practices/Principles*** | ***Freq. Delivery*** | ***Tech. Excellence*** | ***Simplicity*** | ***Emp. Teams of motivated individuals*** | ***Const. Development Pace*** | ***Acc. Change*** | ***Continual stkhlder comm. and coll.*** | ***Freq. refln and learning*** | ***Striving for cust. satisfaction*** |
| Software Configuration Management | | X | | | | | | | |
| Code Ownership (collective and individual) | | | | X | | | | | |
| Iteration Progress tracking and reporting | | X | | | | | | X | |
| Adherence to coding standards | | X | | | | | | | |
| ***Practices/Principles*** | ***Freq. Delivery*** | ***Tech. Excellence*** | ***Simplicity*** | ***Emp. Teams of motivated individuals*** | ***Const. Development Pace*** | ***Acc. Change*** | ***Continual stkhlder comm. and coll.*** | ***Freq. refln and learning*** | ***Striving for cust. satisfaction*** |

***Acknowledgement***

I thank Vinh To for his help in substantiating the linkages presented in this section.

**Evidence from literature for the substantiated linkages presented in tables B.1 and B.2.**

# Appendix C

As mentioned earlier, our approach to assessing (i) the capability of an organization to support the instantiation of an agile method, and (ii) the effectiveness of the method itself, involves the identification and use of indicators. An indicator is a practice, property pair. Properties are observable characteristics of the people, process, project, and product that are reflective of the environment of the organization. We have identified a partial list of observable properties for the practices that are a part of the OPP framework. These properties were defined by answering the set of five questions discussed previously in Section 3.2. Table C.1 below represents our initial list of observable properties. This list is not complete.

*Table C.1. Observable properties of the people, process, project, and product*

| ***Practices*** | ***Observable properties of the people, process and project for assessing capability*** | ***Observable properties of the product and the process artifacts created for assessing effectiveness*** |
|---|---|---|
| Iterative and Incremental Devt. | 1. Existence | 1. Potentially shippable product at the end of each iteration/ release-cycle |
| Evolutionary Requirements | 1. Existence of JIT gathering/ refinement of details<br>2. Feature/ story-driven decomposition of user needs/ requirements<br>3. Existence of a prioritization scheme – customers determine the priorities at the feature-level and the developers at the story and task level<br>4. Supporting tools | 1. Well defined and prioritized stories and features<br>2. JIT refinement of features and stories<br>3. Features, stories and tasks are validated by continuous feedback<br>4. Features, stories and tasks are of the right level of decomposition |
| Refactoring | | 1. Preserve external behavior<br>2. Maintainability |
| TDD | 1. Existence<br>2. Expertise | |
| Automated test builds | 1. Tool support | |
| Pair Programming | 1. One computer for two programmers<br>2. Only one person can write code at a time<br>3. People willing to share knowledge | 1. Continuous code review<br>2. Knowledge sharing |
| Minimal BRUF, and BDUF | | 1. No full-fledged design documentation |
| JIT | | 1. JIT refinement of features and stories |
| Self-Organizing teams | 1. Management is not involved in team building | |
| Working environment reflective of agile philosophy | 1. Open work spaces – war room<br>2. White boards | |
| Continuous Delivery | | 1. Small releases at regular intervals |
| Constant Velocity | 1. Use of tools such as burn-up, burn-down and velocity charts | 1. Constant velocity over multiple iterations |
| Collocated customers | 1. Customer is available onsite | |
| Continuous Feedback | 1. Stakeholders provide feedback regularly | |
| Prioritization | | |
| Daily Progress Tracking Meetings | 1. Existence | 1. Meetings do not exceed 10 to 15 minutes |
| Agile Documentation | 1. Use of index cards to record features and stories<br>2. Use of a software system to manage documentation<br>3. Just enough documentation | |
| Agile Project Estimation | 1. Customers and the development team are involved in the estimation process | |
| Frequent face-to-face communication | 1. Stakeholders meet at regular intervals | |
| Retrospective meetings | 1. At the end of each iteration, the stakeholders meet to discuss what could be changed to make the process more effective | |
| Client-driven iterations | 1. Existence | |
| Smaller and frequent releases | 1. Existence<br>2. Length of each iteration<br>3. Length of each release-cycle | |
| Appropriate composition of expertise | 1. 30% Cockburn levels 2 and 3 people | |
| Software Configuration Management | 1. Existence of software tools | |
| Code Ownership (collective and individual) | | |
| Iteration Progress tracking and reporting | 1. Use of burn-up and burn-down charts | |
| Adherence to coding standards | 1. Existence | |
| ***Practices*** | ***Observable properties of the people, process and project for assessing capability*** | ***Observable properties of the product and the process artifacts created for assessing effectiveness*** |